# Unwinding NFTs in the shadow of IP law

Runhua Wang[1] | Jyh-An Lee[2,3] | Jingwen Liu[2]

[1]University of Science and Technology Beijing, Beijing, China

[2]The Chinese University of Hong Kong Faculty of Law, Shatin, Hong Kong

[3]Washington University in St. Louis School of Law, St. Louis, Missouri, USA

**Correspondence**
Jyh-An Lee, The Chinese University of Hong Kong Faculty of Law, Shatin, Hong Kong.
Email: jalee@cuhk.edu.hk

**Abstract**

Amid the surge of intellectual property (IP) disputes surrounding non-fungible tokens (NFTs), some scholars have advocated for the application of personal property or sales law to regulate NFT minting and transactions, contending that IP laws unduly hinder the development of the NFT market. This Article counters these proposals and argues that the existing IP system stands as the most suitable regulatory framework for governing the evolving NFT market. Compared to personal property or sales law, IP laws can more effectively address challenges such as tragedies of the commons and anticommons in the NFT market. NFT communities have also developed their own norms and licensing agreements upon existing IP laws to regulate shared resources. Moreover, the IP regimes, with both static and dynamic institutional designs, can effectively balance various policy concerns, such as innovation, fair competition, and consumer protection, which alternative proposals struggle to provide.

## 1 | INTRODUCTION

Although Bill Gates has argued that non-fungible tokens (NFTs) are "100% based on greater fool theory" and can hardly improve the world,[1] the NFT market continues to flourish.[2] There are thousands of NFT sales every day.[3] The most expensive NFT was sold for US$91.8 million as of April 2023.[4] Nevertheless, the market is arguably misleading and fragile. Some zealous NFT participants chose to torch physical artworks after minting digital photographs of the original on blockchains as NFTs and offering them for sale, raising questions not only about the legal but also the ethical ramifications of the stunt.[5] Before significant market fluctuations began in early 2022,[6] the majority of NFTs sold for below US$200, and only a small portion of NFTs sold for over US$700.[7] Profits in the secondary market recently shrunk, declining by over 80% in 2022.[8] While some consumers purchase NFTs for digital collections,[9] most NFT buyers do not recoup their investments.[10] In addition, the market continues

---

[1]Mark Bergen, *Bill Gates Blasts Crypto, NFTs as Based on 'Greater-Fool' Theory*, Bloomberg News (June 14, 2022, 5:09 PM), https://www.bloomberg.com/news/articles/2022-06-15/bill-gates-blasts-crypto-nfts-as-based-on-greater-fool-theory#xj4y7vzkg; Ryan Browne, *Bill Gates Says Crypto and NFTs Are '100% Based on Greater Fool Theory,'* CNBC (June 15, 2022, 7:47 PM), https://www.cnbc.com/2022/06/15/bill-gates-says-crypto-and-nfts-are-based-on-greater-fool-theory.html.

[2]Statista Research Department, *Market Cap of Art Blocks NFT Projects Worldwide November 2020–June 2022*, Statista (Apr. 24, 2023), https://www.statista.com/statistics/1291885/art-blocks-nft-market-cap/ ("As of June 30, 2022, the market cap of Art Blocks NFT projects available on the Ethereum blockchain and listed on OpenSea was worth roughly 840 million U.S. dollars.").

[3]*See* Thomas McGovern, *How Many NFTs Are Sold per Day in 2023? [Updated]*, Earth Web (June 23, 2023), https://earthweb.com/how-many-nfts-are-sold-per-day/ ("The most recent data shows that on average, in March 2022, 3200 NFTs are sold per day. In contrast to January where the per day average was 26,000.").

[4]*See* Fang Block, *PAK's NFT Artwork "The Merge" Sells for $91.8 Million*, Penta (Dec. 7, 2021, 6:03 PM), https://www.barrons.com/articles/paks-nft-artwork-the-merge-sells-for-91-8-million-01638918205.

[5]*See, e.g.*, Adam Iscoe, *Burnt Banksy's Inflammatory N.F.T. Not-Art*, New Yorker (May 10, 2021), https://www.newyorker.com/magazine/2021/05/17/burnt-banksys-inflammatory-nft-not-art.

[6]*See* Statista Research Department, *Total Value of Sales Involving a Non-Fungible Token (NFT) in the Art Segment Worldwide Over the Previous 30 Days from April 15, 2021 to July 15, 2023*, Statista (July 20, 2023), https://www.statista.com/statistics/1235263/nft-art-monthly-sales-value/. As of April 15, 2022, 53.6% of NFTs were sold for below $200, and only 17.9% of NFTs were sold for above $700. *Id*.

[7]*See* Kimberly Parker, *Most Artists Are Not Making Money off NFTs and Here Are Some Graphs to Prove It*, Medium (Apr. 19, 2021), https://thatkimparker.medium.com/most-artists-are-not-making-money-off-nfts-and-here-are-some-graphs-to-prove-it-c65718d4a1b8.

[8]*NFT Market Collapse: Reasons, Effects, and Future*, UPYO Blog (Feb. 19, 2023), https://upyo.com/en/post/nft-market-collapse ("An 88% drop in the trading volume on non-fungible token platforms. An 80% drop in the value of the non-fungible tokens.").

[9]*See* Alex Silverman, *Millennials, Not Gen Zers, Are Driving the Recent Physical and Digital Collectibles Boom*, Morning Consult (Apr. 5, 2021, 12:01 AM), https://morningconsult.com/2021/04/05/millennials-nfts-collectibles/.

[10]*See* Hannah Murphy & Joshua Oliver, *How NFTs Became a $40bn Market in 2021*, Fin. Times (Dec. 30, 2021), https://www.ft.com/content/e95f5ac2-0476-41f4-abd4-8a99faa7737d.

---

to struggle with legal uncertainties related to fraud, money laundering, intellectual property (IP) infringement, and unfair competition.[11]

The opportunities and uncertainties inherent in the NFT and other crypto industries have garnered attention from the United States Congress. In March 2021, Congressmen Patrick McHenry (R-NC) and Stephen Lynch (D-MA) introduced the Eliminate Barriers to Innovation Act, which passed in the House.[12] The bill sought to require the Securities and Exchange Commission (SEC) and the Commodity Futures Trading Commission (CFTC) to establish a joint working group to analyze and provide recommendations regarding the regulation of digital assets, including NFTs.[13] On October 5, 2021, McHenry introduced the Clarity for Digital Tokens Act and called for a "safe harbor" for digital asset start-up projects to further promote these industries.[14] Senator Michael Bennet (D-CO) and Congressman Peter Welch (D-VT) introduced bills requiring digital platforms to protect consumers and promote competition.[15] On June 9, 2022, Senators Thom Tillis (R-NC) and Patrick Leahy (D-VT) drafted a letter requesting that the United States Patent and Trademark Office (USPTO) and the United States Copyright Office jointly study the current and potential applications of NFTs and their respective IP and IP-related challenges.[16] Nonetheless, to date, Congress has not introduced any legislation to effectively govern the NFT market, partly because of its struggle to comprehend the rapidly changing market.

Compared to the financial law approach, where legislators have started exploring legislation,[17] the IP community lags behind in terms of reaching a consensus. Some commentators believe in the effectiveness of IP laws in regulating NFT-related activities, while expressing concerns about their potential improper applications.[18] Others have differentiated NFTs from more traditional forms of IP and defended some applications of NFTs as noninfringing behaviors.[19] There have also been proposals to apply property and sales law to govern NFTs with the intention to preempt the application of IP laws altogether.[20] This Article aims to rebut efforts to forestall the application of IP laws to NFTs. We argue that such proposals reveal a fundamental misunderstanding and belittlement of IP laws.[21] IP laws provide a holistic policy framework that considers the intricate interplay of technological, economic, and legal features of the information economy. Such a framework cannot be substituted by financial regulations or other categories of private law.[22] It is essential to recognize that IP doctrines are not rigidly inflexible. They are purposefully designed to strike a delicate balance between a wide array of public and private interests.[23] For example, IP laws can directly or indirectly address crucial concerns related to consumer and environmental protection, which are the aspects often overlooked by financial regulations or other private laws proposed to govern NFT issues.[24]

This Article proceeds as follows. Section 2 explores the legal nature of NFTs, embracing their technological and economic features. It evaluates proposals to apply property and sales law to NFT minting and transactions, analyzes the property and contract attributes of NFTs,

---

[11] *See* Andrea Kramer, Alexandra Scheibe & Rachel Rosen, *Applying Anti-Money Laundering Practices to NFT Market*, LAW360 (Aug. 15, 2022, 6:28 PM), https://www.law360.com/articles/1521137/applying-anti-money-laundering-practices-to-nft-market. *See also* Felicia J. Boyd, Abida Chaudri & Jamie Brazier, *Hermès' Challenge of 'MetaBirkin' NFTs Foretells Future Trademark Litigation Trends*, IPWATCHDOG (June 30, 2022, 7:15 AM), https://www.ipwatchdog.com/2022/06/30/hermes-challenge-metabirkin-nfts-foretells-future-trademark-litigation-trends/id=149916/. *See* Nir Kshetri, *Scams, Frauds, and Crimes in the Nonfungible Token Market*, 55 COMPUT. 60, 61–63 (explaining the types of scams and frauds that could occur in the NFT market). *See also* Christopher Boone & Melissa Landau Steinman, *"Are You Guys Into Crypto????": Celebrities Promoting Cryptocurrencies Become Class Action Targets*, JD SUPRA (Jan. 19, 2022), https://www.jdsupra.com/legalnews/are-you-guys-into-crypto-celebrities-8613768/ (discussing the class action brought against celebrity endorsers promoting the EMAX Tokens based on a California unfair competition law violation cause of action); *in re* Ethereummax Investor Litigation, No. 2:22-cv-00163, 2023 U.S. Dist. LEXIS 180300 (C.D. Cal. Oct., 3, 2022)).

[12] *See* H.R. 1602, 117th Cong. (2021).

[13] *Id.*

[14] *See* H.R. 5496, 117th Cong. (2021).

[15] *See* S. 4201, 117th Cong. (2022); H.R. 7858, 117th Cong. (2022).

[16] *See* Caroline Rimmer, *Senators Ask USPTO and US Copyright Office to Conduct NFT Study, with a Focus on IP Issues*, JD SUPRA (July 11, 2022), https://www.jdsupra.com/legalnews/senators-ask-uspto-and-us-copyright-2608333/.

[17] *See* Press Release, Sen. Kristen Gillibrand, Lummis, Gillibrand Introduce Landmark Legislation to Create Regulatory Framework for Digital Assets (June 7, 2022), https://www.gillibrand.senate.gov/news/press/release/-lummis-gillibrand-introduce-landmark-legislation-to-create-regulatory-framework-for-digital-assets (targeting financial innovation related to NFTs, such as digital assets, blockchain technology, and cryptocurrencies).

[18] *See, e.g.*, NFTJedi, *Hermes v. Mason Rothschild: Dispute over Unauthorized Depictions of Hermes "Birkin" Bags in NFTs Sold by Rothschild*, NOUNFT (Dec. 12, 2021), https://nounft.com/2021/12/12/hermes-v-mason-rothschild-dispute-over-unauthorized-depictions-of-hermes-birkin-bags-in-nfts-sold-by-rothschild/; Andy Ramos, *The Metaverse, NFTs and IP Rights: to Regulate or Not to Regulate?*, WIPO MAGAZINE (June 2022), https://www.wipo.int/wipo_magazine/en/2022/02/article_0002.html ("[L]ittle debate on the application and validity of the current [IP laws] to NFTs and the metaverse."); Emily Dieli, Note, *Tarantino v. Miramax: The Rise of NFTs and Their Copyright Implications*, 2022 B.C. INTELL. PROP. & TECH. F. 1, https://lira.bc.edu/work/ns/d13d6666-6c67-4fd7-9cdd-898bced9a5da.

[19] *See, e.g.*, Jason Blackstone, *NFTs and IP Law: An Overview for Buyers and Sellers*, IPWATCHDOG (Mar. 12, 2022, 12:15 PM), https://www.ipwatchdog.com/2022/03/12/nfts-ip-law-overview-buyers-sellers/id=147405/ ("the sale of NFTs wouldn't appear to threaten the trademark rights of an IP owner."); Andres Guadamuz, *Non-Fungible Tokens (NFTs) and Copyright*, WIPO MAGAZINE (Dec. 2021), https://www.wipo.int/wipo_magazine/en/2021/04/article_0007.html ("[M]ost tokens are not the work itself, but metadata of the work, and making such a token may not infringe copyright.").

[20] *See* Joshua A.T. Fairfield, *Tokenized: The Law of Non-Fungible Tokens and Unique Digital Property*, 97 IND. L.J. 1261, 1290–1300 (2022) (proposing to govern NFT issues under property law and sales law, preempting IP laws).

[21] It is not controversial that IP aligns with property theory and property law in general, but the details are divergent in the development of IP laws and policies. *See* Dave Fagundes, *Introduction: What's Real? IP Through the Lens of Property Theory*, 57 HOUS. L. REV. 261, 261–64 (2019). *See also* Kali Murray, *What Is Owed: Obligation's Relevance in Property and Intellectual Property Theory*, 2 TEX. A&M J. REAL PROP. L. 275, 276 (2015) ("[A]dopting the framework of property law does not restrict us only to a reductive debate over the scope of the intellectual property owners' rights, but permits the consideration of social, political, and competitive impacts of designating certain information as intellectual property.").

[22] Financial regulations protect investors rather than product consumers. Kristen Busch introduced a more comprehensive policy proposal for NFT governance that not only encompasses financial regulations but also addresses concerns related to consumer protection, intellectual property, and energy and environment. KRISTEN E. BUSCH, CONG. RSCH. SERV., R47189, NON-FUNGIBLE TOKENS (NFTS) (2022). *See also* Murray, *supra* note 21, at 276. ("[A]dopting the framework of property law does not restrict us only to a reductive debate over the scope of the intellectual property owners' rights, but permits the consideration of social, political, and competitive impacts of designating certain information as intellectual property.").

[23] *See* Runhua Wang, *New Private Law? Intellectual Property "Common-Law Precedents" in China*, 89 UMKC L. REV. 353, 374–79 (2020) (discussing the private law characteristics of IP laws in comparison with their public law characteristics). *See also* Stewart v. Abend, 495 U.S. 207, 236 (1990) (explaining that the fair use doctrine "permits courts to avoid rigid application of the copyright statute when, on occasion, it would stifle the very creativity which that law is designed to foster"); Google v. Oracle, Inc., 141 S. Ct. 1183, 1197 (2021) (indicating that Copyright Act's fair use provision "set[s] forth general principles, the application of which requires judicial balancing, depending upon relevant circumstances.").

[24] *See, e.g.*, 35 U.S.C. § 292(a); Kate Nuehring, Note, *Our Generation's Sputnik Moment: Comparing the United States' Green Technology Pilot Program to Green Patent Programs Abroad*, 9 NW. J. TECH. & INTELL. PROP. 609, 610–11 (2011); Robert G. Bone, *Hunting Goodwill: A History of the Concept of Goodwill in Trademark Law*, 86 B.U. L. REV. 547, 549 (2006); Rebecca Tushnet, *Registering Disagreement: Registration in Modern American Trademark Law*, 130 HARV. L. REV. 867, 874 (2017). *See generally* Bronwyn H. Hall & Christian Helmers, *The Role of Patent Protection in (Clean/Green) Technology Transfer*, 26 SANTA CLARA COMPUT. & HIGH TECH. L.J. 487 (2010).

and identifies scarcity as an exaggerated technical aspect of NFTs, which has attracted many NFT participants but then set them struggled with property or sales law issues. We also illustrate how NFTs' unique economic feature of rivalry supplements the excludability enabled by IP laws, thereby promoting innovation and advancing the goals of IP policies. Section 3 draws on the classical theories of the tragedy of the commons and the tragedy of the anticommons to explain inefficiency problems in the NFT market that arise without the intervention of IP laws. Section 4 reviews the current practice of IP licensing in the NFT market and provides evidence of the importance of applying IP laws to NFTs. Section 5 proposes an IP-based framework with both static and dynamic institutional designs to regulate the NFT market. Section 6 offers recommendations for the use of IP strategies (i.e., filing, enforcement, and licensing) for artists, designers, and other IP holders. Section 7 concludes.

## 2 | THE LEGAL NATURE OF NFTs

In 2021, following Christie's eye-catching auction of Beeple's "Everydays: The First 5,000 Days" for US$69 million,[25] NFTs became one of the most lucrative markets in the digital era.[26] From Nike's offering of NFT sneakers through RTFKT Studios,[27] to Lamborghini's first-ever NFT project, "Space Time Memory,"[28] to J.P. Morgan's launch of its virtual branch in the decentralized Metaverse,[29] the "buzz" regarding expansion into the digital market was both self-revealing and rapidly spreading.[30] There has also been an accompanying surge in NFT marketplaces.[31] Some buyers perceive NFTs as the next Bitcoin, intending to replicate the success of Bitcoin millionaires who seized the opportunity to invest at an early stage.[32] Some buyers acquire NFTs solely as investments, while others collect them as a gesture of fandom.[33]

As the NFT market continues to grow and policymakers consider regulatory measures, the initial inquiries that demand attention pertain to the nature of NFTs and their potential societal impacts, whether positive or negative. This section employs a legal dispute as an example to elucidate challenges inherent in the NFT market. It proceeds to delineate the economic and legal characteristics that define NFTs and underscores their illusory feature of scarcity. All these characteristics reveal the inadequacy of categorizing NFTs as personal property and underline the imperative for a more in-depth exploration of the legal nature of NFTs.[34]

### 2.1 | A limited definition of NFTs

Technically, an NFT is a data unit stored on a digital ledger called a "blockchain"[35] pointing to the ownership and authenticity of a specific asset.[36] This definition has two prongs: first, an NFT is a unique dataset containing a tokenID and a blockchain address[37]; second, an

---

[25]Taylor Locke, *Mark Cuban Is Bullish on NFTs, and Now They're Mainstream After a $69 Million Auction at Christie's*, CNBC (Mar. 11, 2021, 10:16 AM), https://www.cnbc.com/2021/02/18/christies-to-auction-beeple-nft-art-and-will-accept-ether-as-payment.html.

[26]*See, e.g.*, Brian L. Frye, *How to Sell NFTs Without Really Trying*, 13 Harv. J. Sports & Ent. L. 113, 113 (2022) ("Suddenly last summer, the internet went nuts for 'non-fungible tokens' or 'NFTs.' In a matter of months, NFT sales swelled from a sleepy slough of the blockchain to a thundering cataract that shows no sign of slaking."); Carol R. Goforth, *How Nifty! But Are NFTs Securities, Commodities, or Something Else?*, 90 UMKC L. Rev. 775, 775 (2022) ("Suddenly, NFTs seemed to be on everyone's mind."); Mark Conrad, *Non-Fungible Tokens, Sports, and Intellectual Property Law Issues: A Case Study Applying Copyright, Trademark, and Right of Publicity Law to a Non-traditional Ownership Vehicle*, 32 J. Legal Aspects Sport 132, 132 (2022) ("NFTs have become a hot commodity among collectors and investors, in some cases selling for millions of dollars."); Daniel S. Cohen et al., *The Coming Blockchain Revolution in Consumption of Digital Art and Music: The Thinking Lawyer's Guide to Non-fungible Tokens (NFTs)*, 26 Cyberspace Lawyer 1, 1 (2021) ("[W]hen Beeple's digital art piece 'Everydays: The First 5000 Days' sold at Christie's for US $69 million, the NFT mania truly began."); An P. Doan et al., *NFTs: Key U.S. Legal Considerations for an Emerging Asset Class*, 24 J. Tax'n Invs. 63, 63 (2021) ("Although the technology that makes NFTs possible has been around for several years, NFTs have very much emerged into public consciousness in 2021."); Jonathan Emmanuel, Gavin Punia & Simi Khagram, *Non-Fungible Tokens: What's All the Fuss?*, 39 Westlaw J. Comput. & Internet 1, 4 (2021) ("Non-fungible tokens (NFTs) have been around for many years but have recently gained huge traction."); Anthony J. Dreyer & David M. Lamb, *Can I Mint an NFT with That?: Avoiding Right of Publicity and Trademark Litigation Risks in the Brave New World of NFTs*, 28 Westlaw J. Intell. Prop. 1, 3 (2021) ("Almost overnight, non-fungible tokens (NFTs) have become a sensation.").

[27]Khristopher J. Brooks, *Nike's New NFT Sneakers Selling for More Than $100,000*, CBS News: MoneyWatch (Apr. 28, 2022, 3:22 PM), https://www.cbsnews.com/news/nike-cryptokicks-nft-blockchain-metaverse-rtfkt. Reportedly, Nike has also filed a patent for tokenized shoes branded "CryptoKicks," *see* Oscar Holland, *How NFTs Are Fueling a Digital Art Boom*, CNN (Mar. 10, 2022, 6:42 AM), https://edition.cnn.com/style/article/nft-digital-art-boom/index.html.

[28]*The NFT Space Time Memory*, Lamborghini, https://nft.lamborghini.com (last visited Aug. 24, 2023).

[29]Christine Moy & Adit Gadgil, *Opportunities in the Metaverse: How Businesses Can Explore the Metaverse And Navigate the Hype vs. Reality*, J.P. Morgan (2022), https://www.jpmorgan.com/content/dam/jpm/treasury-services/documents/opportunities-in-the-metaverse.pdf.

[30]Sinéad Carew, *NFT-Related Stocks Gyrate Wildly as Digital Asset Buzz Grows*, Reuters (Mar. 18, 2021; 1:57 PM), https://www.reuters.com/business/nft-related-stocks-draw-attention-digital-asset-buzz-grows-2021-03-18/. Tim Davis, Carey Oven & Robert Massey, *What's All the Buzz about the Metaverse?*, Deloitte (Mar. 2022), https://www2.deloitte.com/us/en/pages/center-for-board-effectiveness/articles/whats-all-the-buzz-about-the-metaverse.html; Arushi Chawla, *NFT: Creating Buzz in Digital Ecosystem*, Counterpoint, (Jun. 16, 2021), https://www.counterpointresearch.com/zh-hans/insights/nft-creating-buzz-in-digital-ecosystem/.

[31]Cryptowisser, a website that lists and rates NFT marketplaces, currently has more than fifty entries and is updated on a weekly basis. Cryptowisser, https://www.cryptowisser.com/nft-marketplaces/ (last visited Aug. 24, 2023). NFT marketplaces act as an intermediary between the minter and purchaser in an NFT transaction in addition to other technological intermediaries. Eduard Banulescu & Maria Petrova, *OpenSea Review: Everything You Need to Know*, Be[in]Crypto (Feb. 1, 2023, 6:30 PM), https://beincrypto.com/learn/opensea-review/. *See* Anthony J. Carbone et al., *Treasury Suggests Limiting Definition of Digital Asset Broker*, Mondaq (Mar. 11, 2022), https://www.mondaq.com/unitedstates/fin-tech/1171128/treasury-suggests-limiting-definition-of-digital-asset-broker ("[T]he transaction involves multiple other intermediaries: the community that runs and maintains the Ethereum protocol, the miner who validates the transaction and causes it to be stored on the Ethereum blockchain, and the creator of the digital wallet software, MetaMask, which stores the private keys controlling Person X's ETH.").

[32]*See, e.g.*, Cohen et al., *supra* note 26, at 1.

[33]*Id* ("NFT purchasers often are collectors who view NFTs as a way to support their favorite artists, actors, musicians, and athletes.").

[34]*See* Juliet M. Moringiello & Christopher K. Odinet, *The Property Law of Tokens*, 74 Fla. L. Rev. 607, 611 (2022).

[35]Frye, *supra* note 26, at 113 ("An NFT is just an encrypted unit of data stored on a digital ledger."); Conrad, *supra* note 26, at 133 ("An NFT is a data unit stored on a digital ledger (or digital wallet), known as a blockchain"); Emmanuel et al., *supra* note 26 ("An NFT is a one-of-a-kind digital token created (or "minted") and recorded on a digital ledger, called a blockchain."); Cohen et al., *supra* note 26, at 1 ("NFTs are non-fungible tokens issued on a distributed ledger such as a blockchain.").

[36]Dreyer & Lamb, *supra* note 26, at 3 ("At a high level, an NFT is a unique digital certificate stored on a blockchain that conveys certain limited ownership rights to an asset, typically a digital one."); Doan et al., *supra* note 26, at 63 ("In general terms, an NFT is a digital asset, based on computer code and recorded on a blockchain ledger to prove ownership and authenticity of a unique asset.").

[37]Andres Guadamuz, *The Treachery of Images: Non-Fungible Tokens and Copyright*, 16 J. Intell. Prop. L. & Prac. 1367, 1370 (2021) ("The first core element to the NFT is a number known as the tokenID, which is generated upon the creation of the token; the second is the contract address, this is a blockchain address that can be viewed everywhere in the world using a blockchain scanner. The combination of elements contained in the token makes it unique: there can only be one token in the world with the combination of tokenID and contract address.").

NFT is an indication of the ownership of a specific digital (and crypto) asset.[38] Although it shares the traits of immutability, authenticity, and scarcity with other crypto assets, such as Bitcoin and Ether,[39] its indivisibility and uniqueness distinguish itself from them.[40] An NFT is not interchangeable, as each NFT is unique and cannot be replaced by another, thereby constituting its "non-fungible" nature.[41] Each NFT is expected to provide a one-of-a-kind authentication of ownership to the underlying asset.[42]

NFT transactions are executed through smart contracts, which are computer codes that are stored and operated on blockchains.[43] Smart contracts ensure that subsequent NFT recipients do not receive any leaner or broader rights compared with the initial rights issued by minters and embedded in smart contracts.[44] In addition to the technological role played by smart contracts, the use of digital assets linked to NFTs in metaverses is governed by the contractual terms provided by the platforms hosting those assets. According to those terms, platforms may limit NFT owners' access to and use of linked assets. Some platforms may even unilaterally delete such assets or delink them from NFTs according to contract terms.[45]

Therefore, it is essential to understand that the ownership of NFTs does not equate to the ownership of the digital assets represented by those NFTs.[46] The acquisition of an NFT typically does not confer ownership or possession of the underlying asset,[47] nor does it involve the transfer of any IP associated with that asset.[48] Some commentators, therefore, view an NFT transaction as granting a limited nonexclusive license for displaying and using the underlying artwork.[49] In essence, an NFT, in its current form, serves as proof of ownership solely for the NFT itself.[50] The interests and legal risks entwined with NFT ownership hinge upon the legal nature of NFTs, which we delve into in the following sections.

## 2.2 | Ownership problems: Who owns the "quantum" NFT

*Free Holdings Inc. v. McCoy* is a representative dispute concerning the ownership challenges surrounding NFTs.[51] At the center of this dispute is the NFT known as "Quantum," which holds historical significance as the very first NFT ever created.[52] It was minted by the artist Kevin McCoy back in May 2014 on the Namecoin blockchain.[53] The Namecoin blockchain, an early platform for NFTs, employed unique combinations of alphanumeric characters to represent specific identifier for those tokens, allowing them to be recorded, claimed, and traded.[54] Notably, Namecoin's rules mandated periodic renewal of NFTs, with failure to renew resulting in the expiration of the NFT record.[55] Consequently, this expiration opened the door for others to re-register and claim ownership over the now expired NFT.[56] McCoy did not update the record for Quantum, so it expired and reentered the pool in January 2015.[57] Subsequently, in 2021, McCoy reminted the Quantum NFT on the Ethereum blockchain, presumably because the latter was "more modern and stable."[58] In video appearances,

---

[38]*See, e.g.*, Goforth, *supra* note 26, at 777 ("NFTs are cryptoassets"); Lewis et al., *Non-Fungible Tokens and Copyright Law*, 33 INTELL. PROP. & TECH. L.J. 1, 1 (2021) ("An NFT is a cryptographic tool using a suitable blockchain, most commonly Ethereum, to create a unique, non-fungible digital asset."). The categorization of NFTs as digital "assets" has recently been recognized by courts in different jurisdictions including the United Kingdom, Singapore, and China. *See* Osbourne v.Ozone, [2022] EWHC 1021 (Comm) (appeal taken from Eng.); Janesh S/O Rajkumar v. Unknown Person, HC/OC 41/2022 (Sing.); Shenzhen Qice Diechu Wenhua Chuangyi Youxian Gongsi Su Hangzhou Yuan Yuzhou Keji Youxian Gongsi (深圳奇策迪出文化創意有限公司訴杭州原与宙科技有限公司) [Shenzhen Qice Diechu Cultural Creativity Co., Ltd. v. Hangzhou Yuanyuzhou Technology Co., Ltd.] (Hangzhou Internet Ct. 2022) (China).
[39]*See, e.g.*, Cohen et al., *supra* note 26 ("[NFTs are] similar to cryptocurrencies like bitcoin in that they can be identified individually and are authenticated through a decentralized system of nodes via a consensus protocol.").
[40]Lewis et al., *supra* note 38, at 18 ("[T]hey differ from cryptocurrencies in that they are each unique, indivisible, and 'non-fungible.'"); Doan et al., *supra* note 26 (noting that the non-fungible nature "distinguishes an NFT from other digital assets.").
[41]To put it more specifically, "the holder of one NFT cannot necessarily obtain equal value by exchanging a token for another NFT within the same ecosystem." *See* James E. Bedar, Patrick Gilman & David Rosenthal, *Tokenization: The Revolution Is Now*, 27 WESTLAW J. DERIVATIVES 1, 11 (2021). *See also* Emmanuel et al., *supra* note 26 (noting that NFTs are "non-fungible," which means unique and not interchangeable because each token comprises the unique data of code and other metadata to be distinguished from other tokens running based on blockchains); Goforth, *supra* note 26, at 777 ("'Non-fungible' means that each NFT is unique, encoded onto the underlying blockchain with certain metadata that makes it different from every other token, even if the underlying work of art or other asset is the same.").
[42]Bedar et al., *supra* note 41, at 3 ("NFTs have been used to transform collectibles—for example, a digital work of art—into one-of-a-kind, verifiable assets that can be traced, authenticated, and sold/traded on-chain.").
[43]Rebecca Carroll, Note, *NFTs: The Latest Technology Challenging Copyright's Law's Relevance within a Decentralized System*, 32 FORDHAM INTELL. PROP. MEDIA & ENT. L.J. 979, 988 (2022).
[44]*See DAO-Enabled NFT Platform*, LEEWAYHERTZ, https://www.leewayhertz.com/dao-enabled-nft-platform/ (last visited Aug. 24, 2023).
[45]*See, e.g.*, João Marinotti, *Can You Truly Own Anything in the Metaverse? A Law Professor Explains How Blockchain and NFTs Don't Protect Virtual Property*, THE CONVERSATION (Apr. 21, 2022, 8:19 AM), https://theconversation.com/can-you-truly-own-anything-in-the-metaverse-a-law-professor-explains-how-blockchains-and-nfts-dont-protect-virtual-property-179267 (using the terms of Sandbox, an NFT and metaverse platform, as an example). *See also* Carroll, *supra* note 43, at 986 (explaining that "whether a buyer holds any ownership or rights to the underlying artwork depends on individual transaction and the market place used for the transaction").
[46]Marinotti, *supra* note 45; Carroll, *supra* note 43, at 986.
[47]Unless the parties agree otherwise. For example, buyers of StockX Vault NFTs, according to the Vault NFT Terms, gains title to both the purchased Vault NFT (i.e., the digital collectible) and the Stored Item (i.e., the physical goods) to which the NFT corresponds. *See* Nike v. StockX, No. 1:22-cv-00983-VEC (S.D.N.Y. July 14, 2022). *See also* Rob Nightingale, *What Do You Actually Own If You Buy an NFT?*, MAKE USE OF (Mar. 29, 2021), https://www.makeuseof.com/what-do-you-actually-own-if-you-buy-an-nft/ ("An NFT is not the digital asset itself. If you buy the NFT for a piece of digital art, the NFT is not the image file. It is only the record of ownership or authenticity that's stored on the blockchain.").
[48]*Id.* (citing Rob Nightingale, *What Do You Actually Own If You Buy an NFT?*, MAKE USE OF (Mar. 29, 2021), https://www.makeuseof.com/what-do-you-actually-own-if-you-buy-an-nft/ [https://perma.cc/9XS4-SWTB]).
[49]Carol Goforth, *What You Should Know Before Buying or Selling an NFT in the US*, COINTELEGRAPH (Apr. 3, 2021), https://cointelegraph.com/news/what-you-should-know-before-buying-or-selling-an-nft-in-the-us.
[50]Frye, *supra* note 26, at 114 ("[T]hey were invented in order to provide indisputable proof of ownership of works of art. But all they can ever prove is ownership of the NFT. . .").
[51]Free Holdings Inc. v. McCoy, No. 1:22-cv-00881-JLC (S.D.N.Y. Mar. 17, 2023).
[52]*See* Kevin McCoy & Jennifer McCoy, *Quantum*, MCCOYSPACE (2023), https://mccoyspace.com/project/125/.
[53]*Id.*
[54]Free Holdings Inc. v. McCoy, No. 1:22-cv-00881-JLC, 2023 WL 2561576, at *1 (S.D.N.Y. Mar. 17, 2023).
[55]The reasons for these requirements are as follows. First, it incentivizes "owners" of words who no longer intend to use them to either let them expire or sell them, thereby decreasing squatting. Second, it ensures that if a word's "owner" loses his or her private keys, the word will eventually be returned to the pool of available words instead of permanently being stuck in limbo. *See* FAQ, NAMECOIN, https://www.namecoin.org/docs/faq/#why-do-names-have-to-be-renewed-regularly (last visited Aug. 24, 2023).
[56]*Id.*
[57]*Free Holdings*, 2023 WL 2561576, at *4.
[58]*See id.* at *14. *See also Sotheby's Sued over Quantum NFT Auction*, LEDGER INSIGHTS (Feb. 4, 2022), https://www.ledgerinsights.com/sothebys-sued-over-quantum-nft-auction/ ("Nowadays, if you're selling an NFT, ideally, it should be on one of the better-known blockchains to make it easily transferable.").

McCoy claimed to have "moved" and "preserved" the Quantum NFT between different blockchains, to which the complete provenance and ownership can be traced.[59] In April 2021, shortly after McCoy's public announcements had garnered its attention, Free Holdings, the plaintiff, swiftly reregistered the Namecoin-based Quantum NFT and claimed ownership. The Ethereum-based Quantum NFT linked to the Namecoin-based Quantum NFT was auctioned at Sotheby's for US$1.472 millions in June 2021.[60] Free Holdings sued McCoy for slander of title, deceptive and unlawful trade practices, and commercial disparagement for the misrepresentation of the ownership and status of the Quantum NFT.[61]

Free Holdings asserted that it acquired the title to the Namecoin-based Quantum NFT and the reminted Ethereum-based Quantum NFT was merely an "authorized print of the original NFT."[62] Free Holdings' core argument rested on McCoy's metadata description of Namecoin when the artist created the Namecoin-based Quantum NFT, as follows:

> I assert title to the file at the URL http://static.mccoyspace.com/gifs/quantum.gif with the creator's public announcement of [its] publishing at the URL https://twitter.com/mccoyspace/status/462320426719641600. The file whose SHA256 hash is d41b8540-cbacdf1467cdc5d17316dcb672c8b43235fa16cde98-e79825b68709a is taken to be the file in question. Title transfers to whoever controls this blockchain entry.[63]

In contrast, McCoy argued that blockchain-based digital art possesses the inherent attribute of immutability, which in turn bestows upon artists ownership and control over their creative works.[64] The court dismissed the complaint on procedural grounds without supporting either of the two arguments.[65] Consequently, while the dispute highlights the technical possibility for artists to create multiple NFTs linking to the same digital artwork on different blockchain platforms such as Namecoin and Ethereum,[66] the legal status of the Namecoin-based Quantum NFT and the ownership of the double-minted NFTs remain uncertain.

## 2.3 | Are NFT properties or contracts?

Numerous participants in the market hold the assumption that NFTs are property because at first glance, they appear to function similarly. This perception is rooted in the belief that NFTs not only facilitate the transfer of ownership of digital assets, but also afford new owners the ability to circumvent legal rights attached to the underlying assets, such as copyrights.[67] The legal scholarship remains divided on the legal nature of NFTs and the rights affiliated with them.

Joshua Fairfield and Lawrence Trautman propose that the legal system should treat NFTs as (digital) personal property because NFTs contain strong property interests, such as enabling parties to buy, sell, or own digital assets similar to personal property.[68] They argue that the sale of an NFT resembles the sale of property, which is the purpose of NFTs' technological design.[69] Therefore, the modalities of an NFT transaction should be similar to the sale of physical artwork, in that buyers do not need to obtain a license from the IP holder when selling the art.[70] Fairfield argues that if the sale of an NFT is viewed as an IP licensing agreement, NFTs would lose the exclusive property characteristics that attract buyers and investors to purchase them.[71] Consequently, applying IP laws to NFT transactions, instead of treating NFTs as personal property, would negatively affect the burgeoning NFT market. To avoid transaction costs stemming from IP licenses, he suggests delinking NFTs from their underlying artworks.[72] A recent ruling by the U.K. High Court of Justice seems to support this view.[73] The court characterized NFTs as properties similar to other crypto assets for the purpose of freezing them through an interim injunction in an NFT fraud case.[74]

The classification of NFTs as property remains a subject of contention among scholars. Edward Lee, while not explicitly rejecting the notion of viewing NFTs as property, underscores the impracticality of dissociating them from their underlying artworks.[75] Lee employs a direct analogy, likening NFTs to library card catalogs, to illustrate the inherent source value that NFTs hold in connection to the artworks they represent.[76] Just as the Dewey Decimal System does not provide an ownership right over the book itself, an NFT does not

---

[59] *Kevin McCoy Quantum*, Sotheby's, https://www.sothebys.com/en/buy/auction/2021/natively-digital-a-curated-nft-sale-2/quantum (last visited June 2, 2022).
[60] *Id.*
[61] Free Holdings sued McCoy, Sotheby's, and the successful bidder at the auction. *Free Holdings*, 2023 WL 2561576.
[62] *Id.* at *6.
[63] *Id.* at *1–2.
[64] Anil Dash, co-creator of the Quantum NFT with Kevin McCoy, commented that, "[t]he only thing we'd wanted to do was ensure that artists could make some money and have control over their work." Anil Dash, *NFTs Weren't Supposed to End Like This*, The Atlantic (Apr. 2, 2021), https://medium.com/the-atlantic/nfts-werent-supposed-to-end-like-this-14f14aff42e1.
[65] *See Free Holdings*, 2023 WL 2561576, at *10–12 (ruling that Free Holdings had failed to establish standing to sue).
[66] Thomas Smith, *Solving the NFT Double Minting Problem with Computer Vision*, Geek Culture (July 5, 2021), https://medium.com/geekculture/solving-the-nft-double-minting-problem-with-computer-vision-c57bbbb4652d.

[67] *See generally* Guadamuz, *supra* note 37.
[68] *See id. See also* Lawrence J. Trautman, *Virtual Art and Non-Fungible Tokens*, 50 Hofstra L. Rev. 361, 412–18 (2022).
[69] Fairfield, *supra* note 20, at 1268, 1292–93.
[70] *Id.* at 1298–1300.
[71] *Id.* at 1290.
[72] *Id.* at 1274, 1295.
[73] Amy Castor, *U.K.'s High Court Ruling on NFTs as 'Property' Has Been Called a Landmark But It May Not Actually Change Much*, Artnet News (May 4, 2022), https://news.artnet.com/market/uk-high-recognizes-nfts-as-property-2108605 (questioning the significance of the ruling).
[74] Lavinia Deborah Osbourne v (1) Persons Unknown (2) Ozone, CL-2022-000110, [2022] EWHC 1021 (March 10, 2022) ("[I]t has been consistently held that crypto assets, are to be treated as located at the place where the owner of them is domiciled. There is no reason at any rate at this stage to treat non fungible tokens in any other way, assuming for present purposes as I do that they are to be treated as property as a matter of English law."). *See also* Castor, *supra* note 73.
[75] *See* Edward Lee, *NFTs as Decentralized Intellectual Property*, 2023 U. Ill. L. Rev. 1049, 1089 (arguing that NFTs represent creations by connecting with digital artworks and are a new form of IP).
[76] *See* Edward Lee, *Episode 3 NFT Myth Busting Buying an NFT is NOT Buying the Art*, YouTube (Oct. 19, 2021), https://www.youtube.com/watch?v=Fy6uh_QRks&ab_channel=nounft.

transfer ownership of an underlying artwork to the owner of the NFT.[77]

Juliet Moringiello and Christopher Odinet invoke property theory to reject any property rights in NFTs.[78] They compare the characteristics of NFTs with those of negotiable instruments, securities, deeds of real property, bills of lading, and other legal tokens and conclude that NFTs lack the tethering effect that other properties have.[79] NFT holders do not control anything substantive. NFTs and the IP over the subject digital contents are independent from each other. Additionally, they note, the content associated with NFTs may be removed in NFT marketplaces.[80] Therefore, Moringiello and Odinet argue that without control over underlying assets, there is no property right to protect in an NFT.[81]

Some commentators argue that NFTs exhibit a contractual nature due to the involvement of smart contracts in facilitating transactions between NFT sellers and buyers.[82] A smart contract comprises a series of code designed to automatically execute an agreement. While we concur that a smart contract should not be equated with a traditional legal contract,[83] we posit that the contractual dimension of NFTs is most vividly illustrated by the terms and conditions stipulated by metaverse platforms where NFTs are transacted. The so-called "ownership" of NFTs is in fact defined by terms of service, meticulously crafted and enforced by the platform companies.[84]

We argue that whether an NFT resembles property or a contract depends on its technological features, which are crucial to clarifying the ownership of a digital asset when it is "hashed" to the blockchain.[85] An NFT may include a pointer to the art, a hash of the art, or both. If the digital asset is stored on a centralized server and the company operating the server ceases to exist, so will the off-chain asset stored there.[86] The legal interests associated with this NFT are consequently subject to contractual or technical restrictions imposed by the server host.[87] Therefore, the off-chain storage of digital assets may entail a more extensive set of contract attributes, thereby imposing rights and liabilities on various NFT stakeholders. In contrast, if the digital asset itself is hashed into the token, which means that it is stored on-chain,[88] the asset will continue to exist even if the company originally hosting the server no longer exists.[89] This also means that the NFT obtains a tethering effect, thereby overcoming the concerns raised by Moringiello and Odinet regarding NFTs' property attributes.[90] The sustainable nature of on-chain storage lends a property-like quality to connected NFTs.[91] However, since it is quite costly to store a large amount of data in tokens, only a few digital assets can be stored on-chain.[92]

Moreover, the property or contractual nature of NFTs depends on the interoperability of a platform. If an NFT can be transferred between platforms with limited control over the NFT and its underlying digital assets, the NFT bears resemblance to property. However, if the NFT can only be displayed or transacted on a single platform, it is more analogous to a contract because the transferability of the token depends on the contractual relations between the platform and the owner. This is the reality of most online games, where players are not able to take items to other games or virtual environments.[93] Currently, many NFT and metaverse platforms are not interoperable, and the asset is not portable.[94] Therefore, the ownership of these assets is subject to the terms of service provided by the individual platform company that hosts the assets.

Finally, ongoing discussions regarding the proprietary nature of NFTs frequently revolve around the concept of exclusivity inherent in property rights. However, it is important to note that the right of exclusivity in property is not boundless. Under certain circumstances, laws permit "free riding" on the property of others, particularly when the social value of such practices exceeds private interests.[95] Fairfield and Trautman raise valid concerns about the transaction costs associated with IP licensing.[96] Nevertheless, they overlook the inherent limitations of IP, such as fair use of copyrighted works and trademarks, which aim to strike a balance between safeguarding creators' interests and fostering upholding the value of free riding. Similarly, Moringiello and Odinet, in their opposition to Fairfield and Trautman's proposals, focus primarily on the exclusivity aspect of the property regime but neglect the deliberate limitations to the exclusivity enshrined in the law.

---

[77]Id.
[78]See Moringiello & Odinet, supra note 34, at 615–27.
[79]See id. at 641–43.
[80]See id. at 634–38 (reminding that the IP right for NFT holders is to display the artworks and breaking the rules of NFT marketplaces for NFTs or laws may result in removing the minted NFTs' contents). See, e.g., Terms of Use, MINTABLE.APP (Dec. 5, 2023), https://d3luz8cn6n4wh0.cloudfront.net/terms_of_use_04_15_2019.pdf ("The Company, however, reserves the right to remove any User Content from the Service at its sole discretion.").
[81]See Moringiello & Odinet, supra note 34, at 643.
[82]See, e.g., Carroll, supra note 43, at 1001 ("NFTs and the blockchain on which it is recorded allow artists to program a set of contracts with the buyer governing the use of the work. This allows the artist to contract directly with the buyer . . . and define what rights, if any, are transferred with the purchase.").
[83]See also Carla L. Reyes, A Unified Theory of Code-Connected Contracts, 46 J. CORP. L. 981, 988 (2021) ("[w]hat becomes clear from even this brief study of the nature of smart contracts is that the word contract is not used in the legal sense of legally enforceable contract. Rather, smart contracts encompass a far greater range of computer programs running on blockchain technology"). Cf. Carla L. Reyes, Emerging Technology's Language Wars: Smart Contracts, 85 WIS. L. REV. FORWARD 85, 104–06, 119–20 (2023) (explaining lawyers' possible misunderstanding of smart contract because their lack of knowledge in the technology).
[84]See, e.g., Marinotti, supra note 45. See also Carroll, supra note 43, at 988 (indicating that "[t]he creation and sale of an NFT is. . .governed by the terms of services in the various marketplaces. Depending on which market place is chosen, different terms will apply to the NFT"); id. at 1001 ("NFTs are . . . bought and sold through third-party marketplaces with their own terms and conditions for the buyers and sellers to follow"); Genevieve Bell, The Metaverse Is a New Word for an Old Idea, MIT TECH. REV. (Feb. 2, 2022), https://www.technologyreview.com/2022/02/08/1044732/metaverse-history-snow-crash/ (defining a metaverse as "an immersive, rich digital world combining aspects of social media, online gaming, and augmented and virtual reality").
[85]Fairfield, supra note 20, at 1274; Trautman, supra note 68, at 365, 371 (indicating that NFT is "a digital proof of purchase that is recorded on a digital ledger known as a blockchain").
[86]Fairfield, supra note 20, at 1283.
[87]Id.
[88]See, e.g., Carroll, supra note 43, at 986.
[89]Fairfield, supra note 20, at 1283.
[90]See Moringiello & Odinet, supra note 34, at 641–43.
[91]Cf. Guadamuz, supra note 37, at 1371 ("on-chain works. . .can only be exchanged and transferred with other people on the blockchain, so the NFT acts more like true ownership of the work").
[92]Fairfield, supra note 20, at 1283. See also id. at 1371 ("[t]here are not many projects that upload the full work to the blockchain [because] the cost of writing data into blockchain is prohibitively expensive and this is so by design").
[93]Trautman, supra note 68, at 400.
[94]See, e.g., Jyh-An Lee et al., Legal Implicatins of Self-Presence in the Metaverse, 25 MEDIA & ARTS L. REV. 267, 277 (2023); Marinotti, supra note 45.
[95]See Mark A. Lemley, Property, Intellectual Property, and Free Riding, 83 TEX. L. REV. 1031, 1032–33 (2005).
[96]See supra text accompanying notes 71–75.

In summary, the propositions advocating for the treatment of NFTs as personal property while rejecting the application of IP laws to NFT-related activities not only overlook the intangible and contractual attribute of NFTs but also fail to grasp the policy importance of constraining exclusivity within IP doctrines. In section 5, we contend that IP doctrines have effectively mitigated the risk of excessive exclusivity in property rights by implementing various limitations.[97] Accordingly, we propose that NFTs involving IP necessitate the full application of IP laws.

## 2.4 | The myth and reality of uniqueness and scarcity

One of the motivations to treat NFTs as property is that doing so avoids the application of IP laws to NFTs. Based on alleged NFT market needs, Fairfield advocates that it is the scarcity associated with NFTs' property status that attracts consumers to pay royalties in both the short run and the long run. Applying IP laws to NFT transactions would, in his opinion, erode the scarcity of NFTs and potentially harm the burgeoning NFT market.[98] Fairfield's argument is based on the assumption that NFTs have a scarcity value.[99] However, is this assumption accurate and a valid argument to exclude IP laws for NFTs?

While many in the NFT community view NFTs as technical solutions independent of current regulations, most agree that NFTs provide a new mechanism for generating information that promotes innovation and creativity.[100] Information has strong characteristics of public goods, such as non-rivalrousness and non-excludability.[101] First, information is non-rivalrous and cannot be depleted because multiple parties can consume the same piece of information simultaneously.[102] The consumption of certain information does not diminish the amount of the same information available to others.[103] Second, information is inherently non-excludable because in its initial form without any human-constructed protection, it does not prevent others from accessing it, often making it impossible to prevent its dissemination.[104] To prevent underproduction of public goods and the problem of free riding, IP, a legal monopoly granted by law, has made IP-protected information exclusive.[105] In other words, IP-protected information is non-rivalrous, but excludable.

In contrast, NFTs, by virtue of their technical properties, can introduce rivalrousness into the underlying digital assets, rendering them scarce resources.[106] Unlike other crypto assets, each NFT is unique,[107] and this uniqueness is what imparts scarcity to NFTs.[108] This scarcity translates into rivalrousness concerning the underlying artwork. In other words, when a digital asset is tokenized and linked to an NFT, the NFT engenders rivalrousness for the asset by constraining its quantity.[109] In this regard, NFTs bear a resemblance to tangible property rather than IP. Consequently, it is often assumed that, given the minting of the underlying asset is on a blockchain, NFT buyers are guaranteed sole ownership of the asset pointed by the NFT.[110] This assumption might lead to the conclusion that NFT buyers are sufficiently protected without intervention of IP laws.

An increasing number of NFT-related frauds, however, have recently been reported, which challenges the uniqueness and scarcity of NFTs. Infringers may sell NFTs linked to the same digital assets across different platforms.[111] While NFTs are supposed to be unique, anyone can mint the same artwork as multiple NFTs.[112]

Moreover, ownership of an artwork represented by an NFT does not prevent technical or legal replication of the underlying work.[113] Similar to the ownership of a physical artwork, which involves two elements—the personal property and the IP encapsulated in the work—NFTs contain similar elements.[114] Despite the blockchain technology inherent in NFTs providing authenticity of ownership and transactions, it falls short in safeguarding the IP in the underlying digital or physical asset. Crucially, neither the underlying asset, such as a digital artwork, nor the NFT itself may inherently possess scarcity. In such instances, Moringiello and Odinet directly challenge the foundational concept of scarcity in NFTs, denying both the scarcity of the underlying digital artworks and the rivalry of NFTs.[115] Scholars like Andres Guadamuz also argue that the scarcity induced by NFTs is "illusory."[116] Nevertheless, the perceived scarcity characteristic of NFTs, even with certain limitations, undoubtedly impacts their value for creators and IP holders, which is a subject explored further below.

---

[97]*See infra* section 5.
[98]Fairfield, *supra* note 20, at 1300.
[99]*Id.* at 1264.
[100]Guadamuz, *supra* note 37, at 1368.
[101]*See* Jyh-An Lee, *Tripartite Perspective on the Copyright-Sharing Economy in China*, 35 COMP. L. & SEC. REV. 434, 435 (2019). *See also* Christopher S. Yoo, *Copyright and Product Differentiation*, 79 N.Y.U. L. REV. 212, 214–15 (2004) (introducing the public goods' economic feature of nonrivalry to copyrights); Niva Elkin-Koren, *Copyright Policy and the Limits of Freedom of Contract*, 12 BERKELEY TECH. L.J. 93, 100 (1997) ("Information is non-rivalrous in the sense that once a work is created, its use by one user does not detract from the use of the same information by others."); Oren Bracha & Talha Syed, *Beyond the Incentive Access Paradigm? Product Differentiation & Copyright Revisited*, 92 TEX. L. REV. 1841, 1848–50 (2014) (introducing that information goods are nonrivalrous, which may be a "public goods problem" resolved by IP).
[102]However, an argument can be made that the value of the information can be exhausted through trade secrets. Fairfield, *supra* note 20, at 1296.
[103]Lee, *supra* note 101, at 435.
[104]*Id.*
[105]*Id.*
[106]*See, e.g., id.* at 1263–64, 1304; Tonya M. Evans, *Blockchain and the Genesis of Creative Justice to Disintermediate Creativity*, 26 LEWIS & CLARK L. REV. 219, 223–24 (2022); Guadamuz, *supra* note 37, at 1370; Trautman, *supra* note 68, at 364, 392. *See also* Carroll, *supra* note 43, at 981 (2022) (arguing that NFTs introduces "scarcity to the internet for the first time").
[107]Fairfield, *supra* note 20, at 1296.
[108]Trautman, *supra* note 68, at 410–11. *See also* Guadamuz, *supra* note 37, at 1383; Fairfield, *supra* note 20, at 1265. *But see* Carroll, *supra* note 43, at 988–89 ("[W]hile the NFT satisfies the scarcity, it does not create the demand.").
[109]Fairfield, *supra* note 20, at 1274.
[110]*Id.* at 1264. *See also* Evans, *supra* note 106, at 222 (describing NFTs as "a means for providing verifiable and immutable proof of ownership of an asset").
[111]Carroll, *supra* note 43, at 1000.
[112]Guadamuz, *supra* note 37, at 1370.
[113]*Id.* at 1384.
[114]*See* Moringiello & Odinet, *supra* note 34, at 644.
[115]*See id.* at 644–47.
[116]Guadamuz, *supra* note 37, at 1383–84 (following Jake Linford and believing that scarcity is a part of copyright). *See also* Jake Linford, *Copyright and Attention Scarcity*, 42 CARDOZO L. REV. 143, 146 (2020).

## 2.5 | The value of NFTs for IP holders and creativity

The technological and economic power of NFTs has generated unprecedented benefits for IP holders that cannot be provided by the traditional IP regime. NFTs foster creativity by advancing the interests of IP holders in two ways: first, NFTs provide a new mechanism for governing fan activities without damaging the economic interests of IP holders, and second, NFTs provide an effective technical architecture for creators to profit from the resale of their works.

The governance of fan activities has posed challenges for IP holders in many industries.[117] On one hand, fan activities can potentially help IP holders expand their reach by promoting their works.[118] On the other hand, fan activities sometimes trigger infringement concerns when fans use works without a license or in a way the IP holder does not endorse.[119] Despite these challenges, NFTs have provided IP holders with a novel way to interact with their fans and reinforce their fan communities. For example, in the video game sector, NFTs have created incentives for players to become increasingly involved in game playing. Although players do not own IPs over the weapons, armors, and other items in the game, they can transact those items with other players via NFTs.[120] Therefore, NFTs facilitate certain transactions of digital assets, which are not stimulated by the current IP system. Such transactions reinforce social ties between members in the same metaverse community that current IP regimes are unable to support.[121]

NFTs have also provided IP holders with a new means of engaging with their fans or consumers. For example, the National Basketball Association (NBA) launched a line of Top Shot NFTs to sell digital collectibles after having sold physical trading cards to fans for many years.[122] Luxury brands, such as Prada, have introduced NFTs linked to their limited-edition products.[123] NFTs have also been integrated into IP holders' Web 3.0 strategies.[124] A noteworthy example is Louis Vuitton's release of NFTs in conjunction with a new game for consumers.[125] In these business models, NFTs can help consumers smooth the transition from the material world to the digital world.[126]

NFTs can help IP holders attract crypto communities as fans and consumers by incorporating NFT art features into their conventional business models. For instance, Gucci introduced virtual celebrities Janky and Guggimon to promote its NFT line, in addition to its on-site service.[127] NFT collectibles can function as tickets that grant access to the ownership of physical artworks or membership in specific communities.[128] For instance, Tiffany & Co. recently introduced custom-designed CryptoPunks pendants known as "NFTiffs."[129] The purchase of NFTiffs identifies loyal customers and allows them to redeem a physical pendant featuring the CryptoPunk character associated with their NFTs. Artists also use NFTs to promote their art and reach new audiences. For example, the heirs of August Sander, a German photographer from the early 1900s, run galleries that exhibit Sander's work and mint NFTs to appeal to younger and international audiences.[130]

NFTs provide another advantage for creators, which is the ability to capture value gains from future transactions.[131] Through smart contracts on a blockchain linked to NFTs, artists can continuously profit from the resales of their works by creating and selling NFTs.[132] This potential profit incentivizes artists to create and sell their works, even though their ability to profit from resales may be limited when NFTs are transacted on multiple incompatible platforms.[133] This technical feature inherent in NFTs resolves the traditional challenge of enforcing *droit de suite*, or the artist's right to collect royalties from the proceeds of any resale of his or her work, which originated in France.[134] Although *droit de suite* has been adopted in approximately 30 jurisdictions,[135] it has been criticized because of the difficulties related to its enforcement.[136] The smart contracts upon which NFTs are built incorporate *droit de suite*, thereby reducing the cost of enforcement. In addition, the transparent nature of blockchains secures the enforcement and provides artists with unprecedented protection.[137]

---

[117]*See, e.g.*, TIANXIANG HE, COPYRIGHT AND FAN PRODUCTIVITY IN CHINA: A CROSS-JURISDICTIONAL PERSPECTIVE 4–5 (2017).
[118]*See, e.g.*, Sean Leonard, *Celebrating Two Decades of Unlawful Progress: Fan Distribution, Proselytization Commons, and the Explosive Growth of Japanese Animation*, 12 UCLA ENT. L. REV. 189, 201 (2005) (recalling that the Japanese animation was promoted in the United States in the 1980s because of the subtitles created by fans and their distribution).
[119]Madhavi Sunder, *Intellectual Property in Experience*, 117 MICH. L. REV. 197, 213–22, 227 (2018).
[120]Fairfield, *supra* note 20, at 1266.
[121]*Id.* at 1274–77. *See also* Carroll, *supra* note 43, at 989–90 (describing that NFTs facilitate artistic communities); Trautman, *supra* note 68, at 392 ("NFT collecting sites are communities.").
[122]*About NBA Topshot*, NBA TOP SHOT, https://nbatopshot.com/ (last visited Aug. 25, 2023).
[123]Kati Chitrakorn, *Prada Expands Web3 Offer with Product-Linked NFT Drop and Discord Launch*, VOGUE BUSINESS (May 31, 2022), https://www.voguebusiness.com/technology/prada-expands-web3-offer-with-product-linked-nft-drop-and-discord-launch; Zoe Sottile, *Tiffany's Sells Out Custom Cryptopunk "NFTiff" Pendants for $50,000 Each*, CNN (Aug. 7, 2022), https://www.cnn.com/style/article/tiffanys-cryptopunk-nft-pendants-trnd/index.html.
[124]Web 2.0 is the current version of the internet. Evolved from Web 2.0, Web 3.0 has the features of "decentralization, trustlessness, and permissionlessness," constructed by the tools like blockchains, protocols, and cryptocurrencies. *See* Margaret James & Timothy Li, *Web 3.0 Explained, Plus the History of Web 1.0 and 2.0*, INVESTOPEDIA (Oct. 23, 2022), https://www.investopedia.com/web-20-web-30-5208698. *See also* Scott Jeffries, *10 Best Web 3.0 Cryptocurrencies to Buy for 2022*, GOBANKINGRATES (Dec. 12, 2022), https://www.gobankingrates.com/investing/crypto/web-3-0-cryptocurrencies/.
[125]Maghan McDowell & Maliha Shoaib, *Louis Vuitton to Release New NFTs*, VOGUE BUSINESS (Apr. 14, 2022), https://www.voguebusiness.com/technology/louis-vuitton-to-release-new-nfts.
[126]Ferragamo is doing something similar. It provides free in-store minting for its consumers at the Soho shop. Consumers may use the NFTs to film a video to at least share on social media platforms. *See* Rachel Wolfson, *NFTs Become Physical Experiences as Brands Offer In-Store Minting*, COINTELEGRAPH (July 9, 2022), https://cointelegraph.com/news/nfts-become-physical-experiences-as-brands-offer-in-store-minting.
[127]*Superplastic and Gucci Present SUPERGUCCI*, VAULT GUCCI, https://vault.gucci.com/en-US/story/supergucci (last visited Aug. 25, 2023) ("For those holding a SUPERGUCCI Janky NFT during the snapshot, [Guggimon] arrives to their wallets in NFT form, and their homes as a ceramic sculpture, reuniting with his second half.").
[128]*See* Edward Lee, *What Are NFTs and the Top 5 Reasons for All the Hype!*, YOUTUBE (Sept. 15, 2021), https://www.youtube.com/watch?v=sVGcLTf0HsY&t=3s&ab_channel=nounft.
[129]*See* Sottile, *supra* note 123.
[130]*See* Kyle Chayka, *What Can N.F.T.s Do for Dead Artists?*, NEW YORKER (May 14, 2022), https://www.newyorker.com/culture/infinite-scroll/what-can-nfts-do-for-dead-artists.
[131]Fairfield, *supra* note 20, at 1280–81; Carroll, *supra* note 43, at 990.
[132]*See, e.g.*, Fairfield, *supra* note 20, at 1280–81; Guadamuz, *supra* note 37, at 1376–77; Trautman, *supra* note 68, at 408–10, 421–22.
[133]Carroll, *supra* note 43, at 1002; Camille Brown, Note, *Coded Copyright?: How Copyright Enforcement, Remuneration, and Verification Terms in Blockchain-Enhanced Contract Models for Online Art Sales Compare to Their Traditional Counterparts*, 31 S. CAL. INTERDISC. L.J. 617, 622, 635 (2022).
[134]*See, e.g.*, Jun Chen & Danny Friedmann, *Jumping from Mother Monkey to Bored Ape: The Value of NFTs from an Artist's and Intellectual Property Perspective*, 31 ASIA PAC. L. REV. 100, 114–15 (2023); Katreina Eden, *Fine Artists' Resale Royalty Right Should Be Enacted in the United States*, 18 N.Y. INT'L L. REV. 121, 123–25 (2005); Michael B. Reddy, *The Droit de Suite: Why American Fine Artists Should Have a Right to a Resale Royalty*, 15 LOY. L.A. ENT. L. REV. 509, 509–10 (1995); David E. Shipley, Droit De Suite, *Copyright's First Sale Doctrine and Preemption of State Law*, 39 HASTINGS COMMC'NS & ENT L.J. 1, 5 (2017).
[135]Reddy, *supra* note 134, at 510.

While NFTs can complement the objectives of the IP system, they also give rise to IP issues, particularly when the NFT minters or buyers are neither IP holders nor licensees. Many NFT buyers assume, albeit naively, that they possess the right to use, display, or transact the underlying digital assets.[138] However, while some platforms may incorporate copyright licenses or arrangements in their transaction terms, a significant number of NFT transactions currently do not transfer any IP interests.[139] As a result, if the underlying digital assets are protected under existing IP laws, then the IP holder would have a legal basis to claim infringement against the NFT owner for unauthorized use.

As this section has illustrated, NFT transactions are rife with significant uncertainties, such as disputes over NFT and IP ownership and whether a particular minting or transaction constitutes IP infringement.[140] To tackle these uncertainties, the following section utilizes classical theories of the commons and anticommons. The analysis aims to elucidate the risks posed by legal frameworks that disregard IP in the face of potential inefficiencies in the NFT markets.[141]

## 3 | INEFFICIENCY IN THE NFT MARKET

Theories of the commons and anticommons have been widely applied to analyze the inefficiencies associated with property and IP laws.[142] The tragedy of the commons has been viewed as the result of too little private control of shared resources, while the tragedy of the anticommons arises from excessive property and control. In this section, we elucidate how leaving IP laws aside could potentially result in both tragedies of the commons and the anticommons, posing a threat to the flourishing NFT market.

### 3.1 | The tragedy of the commons

The tragedy of the commons, a critical theory that explains inefficiency issues, was first coined by Garrett Hardin.[143] He argues that if resources were free and open, humans would exploit them as much as possible and the consequences of overpopulation would be inevitable.[144] A typical example is unlimited pollution: instead of purifying waste, rational people would discharge it into the commons at low costs.[145] Consequently, without limitations, the Earth would become overloaded with waste.[146]

The rapidly growing NFT market and the surge in NFT prices have led to a tragedy of the commons. Participants in the NFT market have competed to mint NFTs from digital assets with promising value, regardless of whether they own the underlying IP or not.[147] Reflecting on Fairfield's strategy to support NFT participants by ignoring IP laws,[148] this omission contributes to the over-minting of digital resources, which could negatively affect the sustainability of Web 3.0. While space on Web 3.0 may seem boundless, the resources essential for its construction, maintenance, and utilization are not.[149] These resources encompass natural resources, such as gas and energy consumed in NFT projects.[150] Moreover, the absence of IP and the associated value render the worth of NFTs groundless and uncertain.[151]

To address the tragedy of the commons, Hardin advocates for the privatization of resources.[152] Applied to the NFT market, this idea would support regarding NFTs as property because of property's assumed characteristic of scarcity. However, as the previous section discusses, without the associated IP, the actual exclusivity of intangible asset is not guaranteed under this model.[153] This is one of the reasons why IP laws are more suitable than property law in regulating NFTs and the ownership of blocks on blockchains. IP laws

---

[136]*See, e.g.*, Alexander Bussey, Note, *The Incompatibility of* Droit de Suite *with Common Law Theories of Copyright*, 23 FORDHAM INTELL. PROP. MEDIA & ENT. L.J. 1063, 1080 (2013); Michelle Janevicius, Droit de Suite *and Conflicting Priorities: The Unlikely Case for Visual Artists' Resale Royalty Rights in the United States*, 25 DEPAUL J. ART, TECH. & INTELL. PROP. L. 383, 427 (2015). *See also* Lara Mastrangelo, Droit De Suite: *Why the United States Can No Longer Ignore the Global Trend*, 18 CHI.-KENT J. IN'L & COMP. L. 1, 3–4 (2018) (noting that the enforcement of *droit de suite* is not effective in some European countries).

[137]Blockchain holders are very likely to fairly compensate artists, rather than to maximize their economic gains in transactions. This prediction comes from the results of the dictator game in the literature. A series of dictator experiments show that people who have the obsolete power to make offers are systematically more benevolent than homo economicus. *See generally* Daniel Kahneman, *Fairness and the Assumptions of Economics*, 59 J. BUS. S. 285 (1986); Robert Forsythe, *Fairness in Simple Bargaining Experiments*, 6 GAMES & ECON. BEHAV. 347 (1994); Christoph Engel, *Dictator Games: A Meta Study*, 14 EXPERIMENTAL ECON. 583, 583–84 (2011).

[138]Fairfield, *supra* note 20, at 1295, 1298.

[139]*See, e.g.*, Guadamuz, *supra* note 37, at 1376; Trautman, *supra* note 68, at 373. *See also* Carroll, *supra* note 43, at 995 ("[W]hen minting an NFT, the NFT does not automatically prove that the seller is the original creator of the item").

[140]*See, e.g.*, NFTJedi, Adidas v. Nike: *The Race to Win the Sneaker, Athletic Wear Metaverse*, NOUNFT (Dec. 14, 2021), https://nounft.com/2021/12/14/adidas-v-nike-the-race-to-win-the-sneaker-athletic-wear-metaverse/ (This would benefit from a parenthetical).

[141]For an argument in favor of applying property law to NFTs in lieu of intellectual property, *see* Fairfield, *supra* note 20, at 1290–95.

[142]*See, e.g.*, Mary L. Lyndon, *Secrecy and Access in an Innovation Intensive Economy: Reordering Information Privileges in Environmental, Health, and Safety Law*, 78 U. COLO. L. REV. 465, 489–90 (2007).

[143]Garrett Hardin, *The Tragedy of the Commons*, 162 SCI. 1243, 1243 (1968).

[144]*See generally id*.

[145]*See id.* at 1245.

[146]*Id.*

[147]*See, e.g.*, Guadamuz, *supra* note 37, at 1376–77 (reminding that anyone can mint various versions of the same artwork into multiple NFTs and artists divide complete works to mint them into multiple NFTs); NFTJedi, *Caked Apes NFTs Turn Sour*, NOUNFT (Mar. 25, 2022), https://nounft.com/2022/03/25/caked-apes-nfts-turn-sour-nasty-lawsuits-embroil-creators-of-caked-apes-who-sue-each-other-over-ownership-taylor-whitley-taylorwtf-sues-jake-nygard-cake-nygard-clare-maguire-antonius-wiriadjaj/.

[148]*See* Fairfield, *supra* note 20, at 1295–96.

[149]Alex Gomez, *Are NFTs Bad for The Environment? (Not Anymore, Here's Why)*, CYBER SCRILLA, https://cyberscrilla.com/are-nfts-bad-for-the-environment/ (last visited June 19, 2022).

[150]*See* Cheyenne Ligon, *How Misinformation on "Book Twitter" Killed a Literary NFT Project*, COINDESK (Nov 23, 2021, 4:59 AM), https://www.coindesk.com/tech/2021/11/22/how-book-twitter-misinformation-killed-a-literary-nft-project/ (explaining the reasons for the failure of a book project on blockchain with young adults writes, the most critical one of which is the harm to the environment by over-minting NFTs).

[151]Even though there is no direct evidence showing that the drop of the market value of NFTs in 2022 resulted from the expansion of the NFT projects with zero royalties, those projects are squeezing the NFT market. *See* NFTJedi, *"Zero Royalties" Policies on NFT Marketplaces Pose Existential Threat to Web3 and Creators*, NOUNFT (Nov. 11, 2022), https://nounft.com/2022/11/11/zero-royalties-policies-on-nft-marketplaces-pose-existential-threat-to-web3-and-creators/.

[152]*See* Hardin, *supra* note 143, at 1245.

[153]*See supra* section 2.

have evolved to incorporate more adaptable policies for tackling issues like free riding and encompass a broader spectrum of conflicting interests arising from technological advancements that seem more appropriate to distribute limited resources than property law.[154] We, therefore, argue that increasing IP litigation against unauthorized minters may alleviate the tragedy of the commons in Web 3.0. This is because stakeholders, such as minters, marketplaces, and buyers of NFTs, may become more cautious about participating in the NFT market due to the inherent risk of IP infringement. Thus, the prospect of litigation might be able to curb the over-minting of digital resources that contributes to a tragedy of the commons in metaverses.

Elinor Ostrom presents an alternative approach to the tragedy of the commons through collective action within the egalitarian communities.[155] She proposes that resource users can formulate coordinated community strategies to avert the tragedy of the commons.[156] NFT communities have devised their own norms and licensing agreements to regulate shared resources, which is a phenomenon explored in detail in section 4.[157] The observation that these communities are driven by private ordering and licensing terms are constructed upon existing IP laws underscores the foundational role of IP laws in enabling NFT communities to navigate the challenges posed by the tragedy of the commons.

Although both Hardin and Ostrom propose viable solutions to the tragedy of the commons, our analysis above indicates that their applications in the NFT market are more inclined toward IP laws than property law. However, the design of IP laws, just like that of property law, may result in the tragedy of the anticommons, which is explored in the following section.

## 3.2 | The tragedy of the anticommons

The theory of the tragedy of the anticommons was first conceptualized in Michael Heller's seminal article in the *Harvard Law Review*,[158] which mirrors Hardin's "tragedy of the commons."[159] The tragedy of the anticommons is a type of market failure that arises from fragmented property rights and coordination breakdown.[160] It occurs when multiple owners of a piece of private property have the right to exclude each other, leading to ineffective use of the property.[161] Wasteful underuse is an outcome of over-privatization.[162]

The anticommons theory has wide implications for the study of innovation and IP.[163] When multiple parties hold rights over IP and are not making them available for free, the high transaction costs of obtaining permission from each IP holder prevent possible downstream innovators from developing new technologies.[164]

Privatization, which is often considered a solution to the tragedy of the commons, leads to at least three types of anticommons problems in the NFT market. These problems stem from the ambiguity surrounding property rights over NFTs. First, copyright co-ownership increases the transaction costs of copyright clearance.[165] It arises when co-owners slide the same artwork to multiple NFTs. Copyright licensees may mint NFTs against IP holders, or the owners of visual artworks may mint NFTs associated with expired IP.[166] Thus, NFT buyers cannot claim clear ownership of the NFTs because of uncertainties surrounding copyright ownership or the scope of copyright licenses for use of the underlying artwork. This lack of clarity decreases the appeal of NFTs.

The second category of anticommons problems emerges from double minting, which creates uncertainties surrounding the valuation of NFTs. Similarly, anyone in possession of a digital artwork can mint an NFT without linking to an NFT previously minted on the same blockchain and tied to the same artwork.[167] Alternatively, individuals may mint NFTs on multiple blockchains, generating multiple NFTs for the same original artwork.[168] The creation of independent Quantum NFTs on both Namecoin and Ethereum, as previously mentioned, demonstrates cross-blockchain double minting.[169] While artists are generally advised against double minting to maximize their value gains, there are no explicit prohibitions against such practices.[170] Beyond artists, some individuals who have acquired neither NFTs nor the IP associated with popular digital artworks, also engage in double

---

[154] Even though DAO's free riders do not deserve punishment for their absence in governance, deploying the strength of IP enforcement and the scope of IP interests in licensing provides a unique alternative to the DAO's governance. *See* Lemley, *supra* note 95, at 1033 (discussing the conventional idea of anti-free riding held by the IP community). *Cf.* Madhavi Sunder, From Goods to a Good Life: Intellectual Property and Global Justice 8–10 (Yale Univ. Press 2012) (explaining the utilitarian goal of IP laws to improve public goods, so that strong IP to eliminate free-riding on IP is unlikely to happen for public interests, such as innovation).

[155] *See generally* Elinor Ostrom, Governing the Commons: The Evolution of Institutions for Collective Action (2015). Robert Cooter & Thomas Ulen, Law and Economics 140 (6th ed. 2016).

[156] *See generally* Ostrom, *supra* note 155.

[157] *Id.*

[158] Michael A. Heller, *The Tragedy of the Anticommons: Property in the Transition from Marx to Markets*, 111 Harv. L. Rev. 621 (1998).

[159] Hardin, *supra* note 143.

[160] *See generally* Heller, *supra* note 158.

[161] *Id.* at 624.

[162] Michael Heller, *The Tragedy of the Anticommons: A Concise Introduction and Lexicon*, 76 Mod. L. Rev. 6, 11–12 (2013).

[163] *See generally* James M. Buchanan & Yong J. Yoon, *Symmetric Tragedies: Commons and Anticommons*, 43 J.L. & Econ. 1 (2000); Michael Heller & Rebecca S. Eisenberg, *Can Patents Deter Innovation? The Anticommons in Biomedical Research*, 280 Science 698 (1998); Jyh-An Lee, *Copyright Divisibility and the Anticommons*, 32 Am. Int'l L. Rev. 117 (2016).

[164] *See generally* Kevin J. Boudreau & Karim R. Lakhani, *"Open" Disclosure of Innovations, Incentives and Follow-on Reuse: Theory on Processes of Cumulative Innovation and a Field Experiment in Computational Biology*, 44 Rsch. Pol'y 4, 4–19 (2015), https://doi.org/10.1016/j.respol.2014.08.001. *See also* Constance Grady, *How an Author Trademarking the Word "Cocky" Turned the Romance Novel Industry Inside Out*, Vox (July 24, 2018, 1:08 PM), https://www.vox.com/culture/2018/5/15/17339578/cockygate-explained-romance-publishing-faleena-hopkins (reviewing a scandal in publishing that authors are asked to change their book title for a generic term registered as a trademark).

[165] *See, e.g.*, Lee, *supra* note 163, at 124.

[166] *See, e.g.*, Chayka, *supra* note 130 (illustrating how the British Museum provides NFTs in their gift shop for the purpose of commercializing the work of deceased artists, such as the paintings made by Katsushika Hokusai, a nineteenth-century artist).

[167] *Free Holdings Inc.*, No. 1:22-cv-00881.

[168] There are different blockchains to mint NFTs, such as BSV, Flow, Solana, Tezos, Eoslo, and Ethereum.

[169] Amended Complaint, Free Holdings Inc. v. McCoy, No. 1:22-cv-00881 (S.D.N.Y. Aug. 15, 2022). *See supra* section 2.2.

[170] *But see* Eric Ravenscraft, *NFTs Don't Work the Way You Might Think They Do*, Wired (May 12, 2022, 8:00 AM), https://www.wired.com/story/nfts-dont-work-the-way-you-think-they-do/ (taking Twitter as an example, which only accepts NFTs from the Ethereum blockchain for now and may or may not change its policies in the future).

minting, raising IP infringement issues.[171] Correspondingly, in order to obtain an NFT with a valid IP license and avoid infringement risks, NFT buyers bear high search or acquisition costs. These costs underscore a tragedy of the anticommons problem in the NFT market, where blockchain technologies alone cannot offer a solution.[172]

The third type of anticommons problem results from "fractionalized NFTs."[173] Some marketplaces batch mint or allow batch minting of NFTs for digital artworks and then sell the NFTs to one or more buyers.[174] These buyers, or those sharing ownership of the same digital art, can neither exclude each other from further exploitation of the underlying digital art nor claim that they are the owners of the only NFT of the art. Furthermore, on some blockchains, including Ethereum, it is impossible to distinguish possessing an NFT from owning it, which leads to further confusion in the NFT market.[175]

If IP interests become fragmented, collaboration between various stakeholders can easily turn adversarial, especially when the scope or nature of the rights is ambiguous.[176] Over the last 2 years, numerous conflicts have arisen due to the unclear delineation of legal rights associated with NFTs. In one example examined in detail in section 5, film director Quentin Tarantino and the film studio Miramax wrestled over NFTs.[177] NFTs also triggered a conflict between the descendants of Pablo Picasso and the holders of his estate.[178] Moreover, while legal disputes over NFT ownership have not yet arisen among individuals or companies that share IP interests (e.g., co-authors), the economic interests and public exposure amplified by NFTs may lead to competition or disputes for the interests.[179]

Although several community-led decentralized autonomous organizations (DAOs) have emerged to help buyers collectively manage their individual NFTs,[180] it is still unclear how these DAOs can govern NFT transactions efficiently given that they operate horizontally.[181] Thus, anticommons problems stemming from fractionalized NFTs remain unresolved.[182] Additionally, the SEC has cautioned against the issuance of fractionalized NFTs, as they might cross the thin line between collectibles and securities.[183]

In line with Heller's theory, a probable outcome of property fragmentation is the underuse of NFTs and the underlying digital assets. This form of anticommons tragedy could impede the expansion of derivative works, thereby limiting the profits that IP holders stand to gain from NFTs.[184] Simply put, the potential tragedy of the anticommons would hinder the emergence of both original and follow-up creative endeavors.

The anticommons problems discussed in this section underline the pivotal role of IP laws in the context of NFTs. When proponents advocating the application of personal property law to NFTs raise valid concerns regarding potential excessive transaction costs associated with IP transactions, they overlook the prospect of mitigating these costs through a clear delineation of the scope of IP and the pertinent license,[185] along with the social benefits provided by IP laws.[186] IP laws have evolved to effectively address the tragedy of the commons by reducing transaction costs among diverse stakeholders and facilitating subsequent commercialization.[187] The next two sections examine commonly employed IP licensing models in NFT transactions and existing IP doctrines. They elucidate how these models and doctrines contribute to the effective governance of the NFT market.

## 4 | IP LICENSES IN NFT COMMUNITIES

Although some scholars critique the application of IP laws to NFTs, participants in NFT communities, including artists and NFT buyers, have collectively forged norms to align the interests of diverse NFT stakeholders.[188] Numerous NFT platforms have integrated IP licenses into the terms and conditions governing NFT transactions. Among these IP licenses, the most successful business model in the market revolves around licenses permitting derivative works and IP commercialization by NFT owners who, in turn, pay resale royalties to IP holders.[189] While these licensing frameworks showcase institutional diversity within the NFT economy, they also serve a crucial role in upholding community norms.[190] Currently, there are three major types of IP licenses or terms that are adopted by different NFT projects and marketplaces: nonexclusive licenses, exclusive licenses, and waivers of IP, as shown in Table 1.[191]

---

[171]"Quantum" is a word in the public domain, which does not trigger IP infringement. However, other disputes concerning double minting might trigger more IP infringement concerns if the corresponding works are copyrightable, such as paintings, images, videos, memes, tweets, columns, music, or even toilet paper. See, e.g., Yuga Labs, Inc. v. Ripps, No. 2:22-cv-04355-JFW-JEM (C.D. Cal. Dec. 16, 2022); Goforth, supra note 26, at 778–79 ("NFTs can be based on all kinds of things. These can include an image (or collage of images), a video, a highlight, a meme, a tweet, a piece of music or anything else–particularly creations that can be digitized."); Frye, supra note 26, at 113 ("[Y]ou could make an NFT of a text, image, or sound file."); Mitchell Clark, NFTs Are in the Toilet: Charmin Is Selling Toilet Paper-Themed Crypto Art, THE VERGE (Mar. 17, 2021, 10:12 AM), https://www.theverge.com/tldr/2021/3/17/22336115/charmin-nft-toilet-paper-cryptoart-marketing.
[172]Smith, supra note 66.
[173]Adarsh Vijayakumaran, Democratizing NFTs: F-NFTs, DAOs and Securities Law, RICH. J.L. & TECH. (Nov. 11, 2021), https://jolt.richmond.edu/2021/11/11/democratizing-nfts-f-nfts-daos-and-securities-law/#_ftnref14.
[174]See, e.g., Shoe-Minting, STEPN WHITEPAPER, https://whitepaper.stepn.com/game-fi-elements/shoe-minting (last visited Aug. 28, 2023) (allowing to mint each sneaker into shoebox NFTs seven times).
[175]Ravenscraft, supra note 171.
[176]The unclear allocation of IP interests in NFTs generates other social costs as well. That is why Lee urges for the creation of a public registry for NFT owners to claim their licensed rights and to avoid the problems of orphan copyrights and orphan NFTs that "pollute" the NFT market. See Lee, supra note 75, at 1095–96 (introducing the problems of orphan works and orphan NFTs and a corresponding solution of a public registry for NFTs). Abigail Bunce, Note, British Invasion: Importing the United Kingdom's Orphan Works Solution to United States Copyright Law, 108 NW. U. L. REV. 243, 256–57 (2014) (introducing the legal uncertainties on the use of orphan works of which the author's information is not available).
[177]See Complaint, Miramax, LLC v. Tarantino, No. 2:21-cv-08979-FMO-JC (C.D. Cal. Nov. 16, 2021). See also infra section 5.3.1.
[178]Chayka, supra note 130.
[179]See, e.g., Zen Faulkes, Resolving Authorship Disputes by Mediation and Arbitration, 3 RSCH. INTEGRITY & PEER REV. 1, 1–2 (2018) (reminding a legal issue on authorship raised after publication).
[180]See What Is a DAO and How Does it Benefit NFTs?, BINANCE (June 10, 2022), https://www.binance.com/en/blog/nft/what-is-a-dao-and-how-does-it-benefit-nfts-421499824684903992.
[181]A horizontally structured organization is usually inefficient in terms of management, but commentators believe that DAOs with such a structure can break this rule to be efficient. DAOs Explained: What Is a Decentralized Autonomous Organization?, INTELLIPAAT (Sept. 4, 2023), https://intellipaat.com/blog/what-is-dao/ (suggesting that enlargement of the pie and the corresponding great incentives eliminate the efficiency concerns).
[182]See Shardeum Community, What Is a DAO and How Does It Work?, SHARDEUM (Mar. 30, 2022), https://shardeum.org/blog/what-is-a-dao-and-how-does-it-work/ (admitting the inefficiency caused by educating and collecting voters, which results in high transaction costs in reforming agreements rather than technically enforcing agreements).
[183]Sarah Tran, SEC Warns Fractionalized Non-Fungible Tokens Could Be Illegal Amid NFT Mania, FXSTREET (Mar. 26, 2021, 6:46 AM), https://www.fxstreet.com/cryptocurrencies/news/sec-warns-fractionalized-non-fungible-tokens-could-be-illegal-amid-nft-mania-202103260646.
[184]But see Edward Lee, The Bored Ape Business Model: Decentralized Collaboration via Blockchain and NFTs, SSRN (Nov. 15, 2021), https://papers.ssrn.com/sol3/papers.cfm?abstract_id=3963881 (arguing that innovation through creating derivative works is growing to align with NFT transactions).
[185]See infra section 5.1.
[186]See infra section 5.3.
[187]See, e.g., Davis v. Blige, 505 F.3d 90, 98–100 (2d Cir. 2007) (clarifying that a co-author can convey non-exclusive rights to the joint work without the consent of his co-author but needs to share benefits from the transaction to other co-authors).
[188]Guadamuz, supra note 37, at 1383.

Certain NFT projects, such as CryptoKitties, CryptoPunks, and Meebits, embrace the NFT License[192]—an open-source license developed by Dapper Labs Inc., the entity behind CryptoKitties.[193] With the NFT License, NFT buyers are nonexclusively licensed "broad and meaningful rights," including general royalty-free noncommercial use and limited commercial use, with earnings not exceeding US$0.1 million annually from NFTs.[194] Another example of the nonexclusive approach is observed on the popular platform Foundation, permitting artists to mint NFTs and disseminate their artworks across blockchains through nonexclusive licenses.[195] Both licenses, however, restrict the scope of rights granted to licensees, either by prohibiting commercial use or by imposing restrictions on its scale. In contrast, certain nonexclusive and unlimited licenses, like the one provided by Moonbirds, confer "full" commercial rights to NFT owners without specifying a revenue cap or exclude others from deploying the digital artworks associated with the NFTs.[196]

**TABLE 1** Common licenses in the NFT market.

|  | Scope of the licensed use | |
|---|---|---|
|  | **Limited** | **Unlimited** |
| Nonexclusive Licenses | CryptoKitties, CryptoPunks, Meebits, Fanaply, KnownOrigin, Foundation, SuperRare, MakersPlace, and La Collection | MoonBirds |
| Exclusive Licenses or Assignment | Vee Friends | BAYC, MAYC, World of Women, and Forgotten Runes |
| Waiver of IP (CC0) | N/A | A Common Place, Anonymice, Blitmap, Chain Runners, Cryptoadz, CryptoTeddies, Goblintown, Gradis, Loot, mfers, Mirakai, Shields, and Terrarium Club |

Some NFT projects or marketplaces adopt an alternative, exclusive licensing approach.[197] For example, the terms and conditions of the Bored Ape Yacht Club (BAYC) explicitly assure NFT buyers of full rights to the underlying artwork, stating that "[w]hen you purchase an NFT, you own the underlying Bored Ape, the Art, completely."[198] This is a typical example where NFT owners can freely deploy artworks without the extensive paperwork involving platforms, minters, and artists. Similarly, the Forgotten Runes NFT License Agreement provides NFT buyers with an exclusive license for commercial use, with minters receiving blanket royalties amounting to 20% of the revenue generated from the buyers' commercial use of the art or derivative works associated with NFTs exceeding US$5 millions.[199] Another example is the World of Women (WoW) digital ownership assignment, which delicately balances the IP interests of NFT owners

---

[189]Lee, *supra* note 184. *See also* NFTJedi, *Following OpenSea's Defense of Creator Royalties, X2Y2 and Blur Abandon Controversial "Zero Royalties" Policies in Favor of Creator Royalties*, NouNFT (Nov. 18, 2022), https://nounft.com/2022/11/18/following-openseas-defense-of-creator-royalties-x2y2-and-blur-abandon-controversial-zero-royalties-policies-in-favor-of-creator-royalties/ (reporting that after OpenSea decided not to use zero royalties, other NFT marketplaces followed and abandoned their use of zero royalties).

[190]*Cf.* Jyh-An Lee, Nonprofit Organizations and the Intellectual Commons 32–33 (2012) (explaining that "institutional diversity" means that various organizations have their own approaches to address issues in the same movement and how license reinforced the norms in the free and open source software movement).

[191]The Forgotten Runes NFT License Agreement describes its sample license as (1) nonexclusive for noncommercial use by people or entities who do not own the NFTs associated with the art and (2) exclusive for commercial use by NFT buyers. However, considering NFTs' excludability, no one other than the purchaser can use the purchased NFT. Meanwhile, as the terms also make clear that each of its characters is linked to an individual NFT, the purchaser is also the only one with a license to use the underlying art, rendering the license de facto exclusive, despite having indicated otherwise. *See Forgotten Runes NFT License Agreement*, Forgotten Runes, https://www.forgottenrunes.com/posts/tos (last visited Aug. 28, 2023).

[192]Daniel Anthony, *Commercializing NFTs – Generating Value from Digital Assets and Intellectual Property Rights*, JD Supra (Mar. 2, 2022), https://www.jdsupra.com/legalnews/commercializing-nfts-generating-value-1110648/#/terms.

[193]*NFT License*, NFT License (Nov. 5, 2018), https://www.nftlicense.org/.

[194]*Id.*

[195]*See, e.g., Terms of Service*, Foundation, https://foundation.app/terms (last visited Aug. 28, 2023).

[196]*See* NFTJedi, *MoonBirds Owners Get "Full Commercial Art Rights for the Moonbird They Own," Apparently Similar to Bored Ape License*, NouNFT (Apr. 22, 2022), https://nounft.com/2022/04/22/moonbirds-owners-get-full-commercial-art-rights-for-the-moonbird-they-own-apparently-similar-to-bored-ape-license/ ("[Moonbirds's] IP license grants 'full commercial art rights to the Moonbird they own'.").

[197]*Id.* ("[Artists] hereby expressly and affirmatively grants to Foundation, and its Affiliates (as defined below) and its and their successors, a non-exclusive, world-wide, transferable, sublicensable, perpetual, irrevocable, and royalty-free license to (a) reproduce, display, perform, distribute and transmit the [artwork], for the purpose of operating and developing [Foundation], and (b) use and incorporate the [artwork], or derivative works of any of the foregoing, on any marketing materials, and to reproduce, display, perform, display and transmit such marketing materials on any media . . . The foregoing licenses include, without limitation, the express rights to: (i) display or perform the [artwork] on [Foundation], a third party platform, social media posts, blogs, editorials, advertising, market reports, virtual galleries, museums, virtual environments, editorials, or to the public; (ii) index the [artwork] in electronic databases, indexes, and catalogues; and (iii) host, store, distribute, and reproduce one or more copies of such [artwork] within a distributed file keeping system, node cluster, or other database . . . or cause, direct, or solicit others to do so.").

[198]*Terms & Conditions*, BAYC, https://boredapeyachtclub.com/#/terms (last visited Aug. 28, 2023). MAYC applies similar rules. *MAYC Terms & Conditions*, BAYC, https://boredapeyachtclub.com/#/mayc/terms (last visited Aug. 28, 2023). The rights inherently exclude moral rights, which are associated with the civil rights of authors and cannot be transferred or assignable. *Cf.* Anthony, *supra* note 192 (supplementing that "A written assignment is required to perfect the copyright transfer.").

[199]*See supra* note 191 and accompanying text.

and artists. Aside from assigning copyright to NFT owners, the WoW agreement explicitly allows them to use the art associated with the NFT as a trademark or in trademark designs, except when the artists or authors deploy the trademarks themselves.[200] Under exclusive licenses, NFT owners become the primary defenders of IP, with minters who are artists or authors retaining the ability to take action if NFT owners do not.[201] Moreover, under this genre of agreement, artists who mint NFTs are not automatically prohibited from using or promoting their artworks. Artists can continue displaying their works, including on social media platforms, until the NFT owners express disagreement.[202] Nevertheless, not all nonexclusive licenses in the NFT market grant licensees or buyers full rights to exploit the underlying works. For example, Vee Friends offers buyers an exclusive licensing scheme, but confines the license solely to noncommercial use.[203]

To boost the visibility and commercial potential of artworks associated with NFTs, an increasing number of NFT platforms and projects require IP holders to either waive their rights or provide an unlimited license to NFT buyers. Some of these platforms adopt a zero royalty policy, enabling NFT buyers to exploit the artwork freely without seeking permission from the original IP holders.[204] While this policy may raise concerns about fairness to artists and other IP holders,[205] these marketplaces are becoming more competitive, even surpassing the market share of OpenSea, one of the most popular NFT platforms.[206] Other platforms, such as CrypToadz and Nouns,[207] go a step further by requiring copyright holders to donate their IP to the public domain through the Creative Commons Zero (CC0) license.[208] This license allows the public to enjoy the broadest scope in exploiting the underlying IP.[209]

While NFT communities hold diverse opinions on the optimal approach to harnessing the commercial value of artworks, the increasing adoption of IP licenses in NFT marketplaces reveals the potential of IP in aligning various interests among stakeholders. Several implications arise from these evolving licensing arrangements within NFT communities. First, these licensing agreements establish well-defined norms for NFT minting and transactions, thereby mitigating the tragedy of the commons associated with over-minting. This trend validates Ostrom's theory that collective action by community members with different preferences can address the tragedy of the commons.[210] Second, despite the variety of licenses that grant NFT buyers different degrees of rights to exploit the underlying artworks, these licenses contribute to clearly defining the decision-maker among all stakeholders regarding the use of the artworks. Consequently, these licenses can alleviate the tragedy of the anticommons that results from fractionalized NFTs. It is essential to recognize that the aforementioned solutions to both tragedies of the commons and the anticommons are grounded in existing IP laws. Hence, the suggestion of removing IP laws from this market appears not only unrealistic but also disrespectful of stakeholders' preferences. Building upon these insights, the following section proposes a governance model that applies IP laws to NFTs based on policy considerations that extend beyond the commercial interests of artists, NFT buyers, and marketplaces, encompassing other stakeholder interests.

## 5 | APPLYING IP LAWS TO THE NFT MARKET

This section begins by providing a normative perspective on NFT transactions within the framework of a conventional static view of IP. We first propose the incorporation of NFT minting within the existing IP license rules. We then illustrate that while traditional IP doctrines serve the interests of NFT transaction participants, they do not comprehensively address certain policy issues arising in the NFT ecosystem. We argue that a rigid, static application of IP laws under a centralized structure proves inadequate for NFT-related activities conducted through blockchains and Internet architecture, both of which are inherently decentralized.[211] Therefore, a more suitable approach involves static IP rules accompanied by dynamic exceptions, providing greater flexibility and adaptability to the swiftly evolving market and technologies.

---

[200]See WORLD OF WOMEN DIGITAL OWNERSHIP ASSIGNMENT 2. In Art. 6, the agreement addresses that NFT minters do not lose their moral rights in NFT transactions. This agreement only authorizes NFT owners to enjoy the interests of copyrights. It does not transfer any commercial rights regarding trademarks. In Art. 7, it claims that "the Owner shall not use the trademarks, service marks, or proprietary words or symbols of the Creator, to the extent otherwise permitted by applicable law or by written agreement of the Creator." In other words, it is an unlimited copyright license but a limited trademark license. World of Women is not the only platform allowing copyright transfers in NFT transactions. Hup Life offered an option of copyright transfer in NFT transactions and its terms respected the Berne Convention, which is broken now. An End to Our Project, HUP (Nov. 1, 2021), https://hup.life/. See also Guadamuz, supra note 37, at 1373.

[201]WORLD OF WOMEN DIGITAL OWNERSHIP ASSIGNMENT, supra note 200, at Art. 8.

[202]Id. at Art. 7 ("The Creator shall have the right, at its sole discretion, to promote, including through social medias, any public use of the Art by the Owner, unless the Owner informs the Creator otherwise.").

[203]See Vee Friends NFT Terms of Use, VEE FRIENDS, Art, 5, https://veefriends.com/terms-of-use (last visited Nov. 7, 2023).

[204]See NFTJedi, Race to the Bottom: Will NFT Creator Royalties Become a Blur?, NOuNFT (Dec. 7, 2022), https://nounft.com/2022/12/07/race-to-the-bottom-will-nft-creator-royalties-become-a-blur/. ZINU is the first NFT platform offering a "royalty-free" NFT license to allow personal and commercial uses by their NFT buyers. See Andrew Rossow, NFT Licenses: How Can Creators Legally Protect Their IP When They Make NFTs?, NFT NOW (May 2, 2022), https://nftnow.com/guides/nft-licenses-how-can-creators-legally-protect-their-ip-when-they-make-nfts/.

[205]See NFTJedi, Race to the Bottom, supra note 204.

[206]See NFTJedi, supra note 151 ("OpenSea is the last hope for stopping the race to the bottom, but OpenSea is shouldering a disproportionate burden, collecting 91 percent of all creator royalties in the entire market while having less than 50 percent of the market of NFT sales."). Scholars like Edward Lee expressed concerns that the spread of zero royalty policies provides Web 3.0 nothing but a path to the grave because NFTs are losing their function of effectively rewarding artists. Lee, supra note 184. See also NFTJedi, supra note 189.

[207]The Complete Guide to CC0 NFTs – Best CC0 NFT Projects, WAGMI TIPS, https://wagmi.tips/guides/cc0-nfts/ (last visited Aug. 28, 2023).

[208]CC0 "No Rights Reserved," CREATIVE COMMONS, https://creativecommons.org/share-your-work/public-domain/cc0/ (last visited Aug. 28, 2023) ("CC0 empowers yet another choice altogether – the choice to opt out of copyright and database protection, and the exclusive rights automatically granted to creators – the "no rights reserved" alternative to our licenses"). See, e.g., NFTJedi, supra note 196 ("mFers and Nouns use CC0 licenses abandoning IP rights or donating them to the public domain.").

[209]See J. Markezic & M. Bacina, Can't Be Evil? Considering Creative Commons NFT Licenses, BITS OF BLOCKS (Sept. 6, 2022), https://www.bitsofblocks.io/post/can-t-be-evil-considering-creative-commons-nft-licenses.

[210]See Elinor Ostrom, Collective Action and the Evolution of Social Norms, 14 J. ECON. PERSPS. 137, 145–47 (2000). See supra text accompanying note 155–57.

[211]See, e.g., Joseph P. Liu, Owning Digital Copies: Copyright Law and the Incidents of Copy Ownership, 42 WM. & MARY L. REV. 1245, 1333 (2001) (criticizing the limits of copyright law's centralized characteristics).

## 5.1 | A static view of applying IP laws in the NFT market

As demonstrated in section 4, existing IP laws lay the groundwork for NFT communities to harmonize the interests of various stakeholders. However, while this avenue of private ordering has tackled the allocation of rights among different market participants to some extent, certain IP issues persist, particularly on NFT platforms that have not integrated licensing terms. NFT minters who lack ownership of the underlying IP still encounter mounting accusations of IP infringement.[212]

Here we argue that IP is the sine qua non to maximize the value of NFTs. If anyone can mint and transact NFTs without permission from IP holders, NFTs would fail to ensure reasonable compensation for artists.[213] To mitigate the transaction costs and uncertainties associated with the tragedy of the anticommons, we propose that the minting of an NFT should always require authorization from the IP holder of the underlying artwork. In our proposal, courts and legislators should clarify that when an IP holder mints NFTs or permits minting by a third party, he or she grants an implied license to the third-party minter, subsequent buyers, and the marketplace to exploit the artwork unless otherwise indicated by the IP holder. This proposal is justified by the increasing awareness among IP holders that the value of NFTs is inseparable from the IP associated with them.[214] Therefore, the object of an NFT transaction includes both the NFT itself and the license for the underlying IP.

Accordingly, by linking an implied IP license to the NFT minting permitted by the IP holder, NFT buyers not only obtain lines of codes but are also licensed to exploit the underlying IP of the digital artwork.[215] Consequently, when an NFT is transferred to a third party, that party is also licensed to exploit the underlying artwork to the extent permitted by the licensing terms.[216] IP holders can grant certain rights to initial and subsequent NFT buyers and marketplaces, including, but not limited to the right to copy, display, redistribute, and prepare derivative works.[217] Thus, minting NFTs and subsequent transactions do not constitute IP infringement if the initial minting is permitted by the IP holder. Furthermore, we suggest that NFT marketplaces should contemplate the necessity of mandating parties who initially mint NFTs to furnish prima facie evidence demonstrating their status as IP holders or licensees.

In our proposal, because IP licenses are incorporated into NFT transactions, existing IP law can effectively solve many legal uncertainties therein. Below are four examples that show how U.S. copyright laws and their application by courts can resolve copyright issues related to NFT transactions.

1. Under current U.S. copyright law, copyright co-owners have the right to grant nonexclusive licenses for their work without the permission of other co-owners.[218] Accordingly, if a copyright license is incorporated in the NFT transaction, NFT buyers need not worry about copyright infringement risks, even if only one or a few of the copyright holders license the copyright.
2. According to established copyright law, if an exclusive license is not agreed upon by all co-owners of a copyright, it can only be viewed as nonexclusive.[219] Therefore, if in an NFT transaction, not all copyright holders agree to extend an exclusive license when the NFT is minted and sold, NFT buyers can still obtain a nonexclusive license.
3. Under existing copyright law, co-owners of a copyright can legally assign their ownership without the agreement of other co-owners.[220] Therefore, if a co-owner of a copyright assigns his or her share of ownership when an NFT is minted, NFT buyers can still receive partial copyright and share the ownership with other co-authors who do not agree to assign their shares.
4. Per U.S. copyright law, a buyer of a physical artwork is not automatically authorized to create digital copies or display the art online.[221] Accordingly, the buyer does not have any statutory privileges to create NFTs for the art without permission from its copyright holders.

Some commenters argue that applying copyright law to NFTs is problematic due to controversies surrounding the application of the first-sale doctrine in the digital space.[222] Generally, the first-sale doctrine limits the control of copyright holders over copies of their works to the first sale or transfer of those works.[223] Once copies are lawfully sold or transferred, the copyright holders' interest in their works is considered exhausted.[224] However, in the United States, courts have confined the first-sale doctrine to physical copies of protected works,[225] declaring it is inapplicable to digital works. The transmission of digital copies typically involves the reproduction of the work. Thus, applying the first-sale doctrine to digital copies would unduly restrict the copyright holders' right of reproduction.[226] Some courts have reached the same conclusion by interpreting the term "copies" in the U.S. Copyright Act as "material objects."[227] This interpretation aligns with the nature of the transaction. While physical copies, such as books, are typically transacted through sales agreements, digital copies are usually and more appropriately governed by licensing agreements.[228]

---

[212]The defendants could be the authors of the art, the art's collectors, or the buyers of products regarding the art. *See, e.g.*, Complaint, Miramax, LLC v. Tarantino, No. 2:21-cv-08979-FMO-JC (C.D. Cal. Nov. 16, 2021); Wallace Ludel, *Digital Art Company Claims It Is Owed Millions from NFT Sales in Lawsuit Against Artist*, ART NEWSPAPER (Mar. 14, 2022), https://www.theartnewspaper.com/2022/03/14/digiart-danny-casale-coolman-coffeedan-lawsuit-nfts (introducing a dispute between an artist Danny Casale and his licensee in NFT uses on the former's sale of NFTs); B.I.G. v. Yes Snowboards, No. CV-19-01946, 2022 U.S. Dist. LEXIS 99870 (C.D. Cal. June 3, 2022) (declining the prohibition from creating NFTs of derivative works); Roc-A-Fella Records, Inc. v. Againstdamon Dash & Godigital Records, No. 1:21-cv-05411-JPC, 2022 U.S. Dist. LEXIS 114591 (S.D.N.Y. June 27, 2022); Tamarindart v. Husain, No. 1:22-cv-00595, 2022 WL 195808 (S.D.N.Y. May 11, 2022).
[213]Carroll, *supra* note 43, at 1007. *See also* Chayka, *supra* note 130 (suggesting that NFT collectors can be a new form of patrons for artists).
[214]*See supra* section 2.5.
[215]*See, e.g.*, Smart contracts regarding BAYC NFT collections usually embed commercial rights, even though the rights are not performed on blockchains and enforced by technologies as the usual function that smart contracts do. By contrast, Larva Labs' CryptoPunk NFTs prohibit commercial uses by NFT buyers. *See* Lee, *supra* note 184.
[216]*See, e.g.*, Economist, *Bored Ape Yacht Club: The Case for Licensed Commercial Use Rights*, MEDIUM (Sept. 17, 2021), https://medium.com/@deconomist/bored-ape-yacht-club-the-case-for-licensed-commercial-use-rights-b1bbd463d189 (introducing various commercial uses after obtaining the BAYC NFTs, such as to print apes on beer bottles in the commercialization by a brewery).
[217]*See, e.g.*, Terms of Service, FOUNDATION, https://foundation.app/terms (last visited Aug. 28, 2023) (leaving the artists who are NFT creators and hold the rights to "reproduce, prepare derivatives of, distribute, and display or perform [the art]" to decide the scope of license in NFT transactions).
[218]*See* Brownstein v. Lindsay, 742 F.3d 55, 68 (3d Cir. 2014) ("With respect to licensing a joint work, each co-author is entitled to convey non-exclusive rights to the joint work without the consent of his co-author."); Strauss v. Hearst Corp., No. 85 Civ. 10017, 1988 U.S. Dist. LEXIS 1427, at 1837 (S.D.N.Y. Feb. 19, 1988) ("The immunity of a joint owner from suit extends to the situation where one co-owner makes unauthorized use of the contribution of the other."); 17 U.S.C. § 201(a).
[219]*See* Marino v. Usher, 22 F. Supp. 3d 437, 443–44 (E.D. Pa. 2014) (explaining that an exclusive license is a de facto non-exclusive license if the license is not approved by all the co-authors).
[220]*See* Thomson v. Larson, 147 F.3d 195, 199 (2d Cir. 1998) (excluding the risks of copyright infringement if someone transfers his or her rights in a joint work without consent from the co-authors). *See also* Brownstein, 742 F.3d at 69 (confirming the shares of profits to a licensee who licenses a joint work without consent from all the co-authors).
[221]17 U.S.C. § 106(3). *See generally* Liu, *supra* note 211.

We argue that even in jurisdictions outside the United States that apply the first-sale doctrine to digital goods, NFTs should be exempted from this general rule. The first-sale doctrine is incompatible with the technical characteristics of NFTs and blockchains. The transfers of NFTs are processed on blockchains through the execution of smart contracts, which require the parties' assent to copyright licensing terms binding each NFT recipient. Additionally, as mentioned earlier, NFT communities typically allow artists to collect royalties from each NFT resale. This prevailing practice implies that these communities have no interest in copyright exhaustion in NFT transactions.

Compared to copyright law, other IP rules tend to favor IP holders in a more straightforward manner. For instance, statutory language unequivocally puts that the trademark owner's permission is necessary for anyone to use, copy, or display their trademarks.[229] There are no compelling reasons for NFTs or other virtual assets to be exceptions to this rule.[230] Its application can effectively resolve disputes between trademark owners and NFT owners who use trademarks associated with the underlying virtual object without consent.[231]

Similarly, minting or transacting NFTs involving trade secrets without the unanimous consent of all owners of the trade secrets arguably constitutes trade secret misappropriation. Due to the confidential nature of trade secrets, co-owners and exclusive licensees cannot grant a license to a third party without the unanimous consent of all owners or some clear waiver of confidentiality.[232] Failing to do so allows co-owners of trade secrets to claim damages against the transacting parties under the Defend Trade Secrets Act or the inevitable disclosure doctrine.[233] The misappropriation concern exists even if the NFT or underlying work does not disclose the information but reveals its existence.[234] We argue that the transparent nature of smart contracts and other blockchain technologies further enhances the risk of trade secret misappropriation.[235]

IP allocations and royalties stipulated in licensing agreements for NFT transactions are not enforceable if the underlying artworks are in the public domain.[236] An increasing number of NFT artworks are being generated by artificial intelligence (AI)[237] and therefore are not eligible for patent or copyright protection.[238] Licensees or consumers have several legal causes of action (e.g., breach of warranty, unjust enrichment or restitution, fraud, and false advertising) against licensors who collect royalties without valid IPs.[239] This is probably why the AI art generator Artbreeder has allowed anyone to freely mint NFTs for its works, as well as access to those works through a CC0 license.[240]

---

[222]See, e.g., Carroll, supra note 43, at 1001. First-sale doctrine in copyright law allows buyers of physical artworks to lend, resell, and distribute the works. Dieli, supra note 18, at 5–7 (suggesting the inapplicable of the first-sale doctrine for NFTs).
[223]17 U.S.C. § 109(a).
[224]Id.
[225]See Capitol Records, LLC v. ReDigi Inc., 934 F. Supp. 2d 640, 656 (S.D.N.Y. 2013).
[226]See, e.g., MARSHALL A. LEAFFER, UNDERSTANDING COPYRIGHT LAW 331–32 (5th ed. 2010).
[227]See 17 U.S.C. § 101. See, e.g., Redbox Automated Retail, LLC v. Buena Vista Home Ent., Inc., 399 F. Supp. 3d 1018, 1032–33 (C.D. Cal. 2019) ("The first sale doctrine applies to "particular" copies that exist in the material world.").
[228]A policy reason behind this interpretation is that "[d]igital information does not degrade," which allows every recipient of the information to enjoy the full value of the copyrighted work. See SoftMan Prods. Co. v. Adobe Sys., 171 F. Supp. 2d 1075, 1083–88 (C.D. Cal. 2001) (distinguishing selling software copies from licensing the software); U.S. COPYRIGHT OFF., LIBR. OF CONG., DMCA SECTION 104 REPORT 82 (2001); Disney Ent. v. Redbox Automated Retail, LLC, No. CV 17-08655 DDP, 1057–59 (C.D. Cal. Feb. 20, 2018).
[229]15 U.S.C § 1114.
[230]But see Benjamin Stasa, Nike v. StockX Case Highlights Many Unanswered Questions About IP and NFTs, JD SUPRA (Sept. 7, 2022), https://www.jdsupra.com/legalnews/nike-v-stockx-case-highlights-many-9205701/ (arguing that NFTs bring open questions to trademark law by taking the dispute between Nike and StockX as an example).
[231]See, e.g., Nike, Inc. v. StockX, No. 1:22-CV-00983 (S.D.N.Y. Feb. 3, 2022). In this dispute, the sports apparel manufacturer, Nike, sued an NFT marketplace, StockX. On January 18, 2022, StockX launched the StockX Vault NFTs, which were minted on the Ethereum blockchain and associated with a unique physical product held in StockX's custody until the NFT owner "redeems" the NFT in exchange for the associated physical product or some other benefit. Upon redemption, StockX will remove the Vault NFT from circulation by deleting it from the owner's portfolio. Allegedly, StockX's Vault NFT collection comprised nine NFTs, eight of which displayed Nike's marks, associated with Nike products. These NFTs were sold at inflated prices, much higher than the retail price of physical Nike sneakers. Likewise, Nike filed a lawsuit against StockX for trademark infringement, dilution, false designation of origin, unfair competition, etc.
[232]See B.F. Gladding & Co. v. Scientific Anglers, Inc., 245 F.2d 722, 729 (6th Cir. 1957) ("Even without any specific mention in the contract joint trade secrets would be protected against unlawful disclosure by one of the parties."). See also, Jamison v. Olin Corporation-Winchester Div., No. 04-76-KI, 2005 U.S. Dist. LEXIS 49119, at *24 (D. Or. Oct. 4, 2005).
[233]Defend Trade Secrets Act of 2016, Pub. L. No. 114–153, 130 Sat. 376 (2016). See also Runhua Wang, Judicial Reward Allocation for Asymmetric Secrets, 40 PACE L. REV. 226, 232 (2020). See, e.g., PepsiCo, Inc. v. Redmond, 54 F.3d 1262 (7th Cir. 1995).
[234]See Brown, supra note 133, at 637. See also Wang, supra note 233, at 237–41 (categorizing business tricks and secrets per se as negative trade secrets, which are valuable without active deployment in business).
[235]Because transactions through smart contracts and blockchains are transparent, the public can track the source of the information associated with NFTs. If the information is confidential, NFTs exposes their existence, which may attract thieves and provide clues to them.
[236]See Paul J. Heald, Payment Demands for Spurious Copyrights: Four Causes of Action, 1 J. INTELL. PROP. L. 259, 260 (1994). See, e.g., MedImmune, Inc. v. Genentech, Inc., 549 U.S. 118, 144 (2007) ("[I]nvoluntary or coercive nature of the exaction preserve[d] the right to recover the sums paid or to challenge the legality of the claim.") (citing Altvater v. Freeman, 319 U.S. 359, 365 (1943)).
[237]See, e.g., Aiarthouse, Artificial Intelligence Art, OPENSEA, https://opensea.io/collection/artificial-intelligence-art (last visited Aug. 29, 2023); Godfrey Benjamin, Cloned Bored Ape NFT Emerges through Google AI Technology, BLOCKCHAIN NEWS (Aug. 4, 2022), https://blockchain.news/news/cloned-bored-ape-nft-emerges-through-google-ai-technology.
[238]See, e.g., Thaler v. Vidal, No. 21-2347 (Fed. Cir., Aug. 5, 2022) (restating that AI are not qualified inventor under patent law); Thaler v. Perlmutter, No. 22-1564, 2023 U.S. Dist. LEXIS 145823 (D.D.C. Aug. 18, 2023) (rejecting copyright protection for artworks generated only by AI); Adi Robertson, The US Copyright Office Says an AI Can't Copyright Its Art, THE VERGE (Feb. 21, 2022), https://www.theverge.com/2022/2/21/22944335/us-copyright-office-reject-ai-generated-art-recent-entrance-to-paradise ("The US Copyright Office has rejected a request to let an AI copyright a work of art."); Rita Matulionyte & Jyh-An Lee, Copyright in AI-generated Works: Lessons from Recent Developments in Patent Law, 19 SCRIPTED 5, 8–9 (2022) (explaining that "AI-generated works are not easily copyrighted in most countries because such laws require human authorship for copyright protection"). But see Jyh-An Lee, Computer-Generated Works under the CDPA 1988, in ARTIFICIAL INTELLIGENCE AND INTELLECTUAL PROPERTY 177, 186–89 (Jyh-An Lee et al. eds., 2021) (explaining that AI-generated works may be viewed as copyright works under the Copyright, Designs and Patent Act (CDPA) 1988 in the United Kingdom).
[239]Heald, supra note 236, at 262.

## 5.2 | From static to dynamic IP rules

Although the IP rules discussed above can properly align the interests of different stakeholders in most circumstances, they are not a panacea for all issues associated with IP in the fast-changing NFT market. Even with existing license agreements, artists and their exclusive licensees continue to dispute who has the right to mint and allow the subsequent sale of NFTs representing the underlying artworks.[241] For example, in 2021, a New York City gallery operator, Tamarind Art (Tamarind), filed a motion for declaratory judgment against the estate administrators of a deceased Indian artist, Maqbool Fida Husain (MFH), to pursue its rights to create and sell NFTs of one of the artist's most popular works.[242] When purchasing the visual artwork in 2002, Tamarind had been granted an exclusive, royalty-free, worldwide license to "display, market, reproduce, and resell all or any part of the artwork (on all digital and offline media), including all IP in respect thereof."[243] The artist and the gallery entered into a supplementary agreement in 2003, which affirmed that Tamarind and its affiliates owned the copyrights for any artworks they had purchased from MFH, and that the artist and their heirs retained no right therein.[244] Tamarind's affiliate launched a campaign in 2022 to sell NFTs based on the disputed artwork, in response to a cease-and-desist letter from the defendants.

There are also instances where artists mistakenly believe they do not need a license to mint NFTs for works based on another's IP. A notable example is the "MetaBirkins" dispute between Hermès and the artist Mason Rothschild.[245] Since December 2, 2021, Rothschild offered and advertised on different channels a line of 100 Ethereum-based digital collectibles called MetaBirkins, which feature Hermès's iconic fur-covered Birkin handbag.[246] After receiving cease-and-desist letters from Hermès, the marketplace OpenSea removed these NFTs. Rothschild then moved the NFTs to Rarible and opened two more stores in NFT marketplaces.[247] Hermès filed a lawsuit in the Southern District of New York, alleging trademark infringement, dilution, false designation of origin, false descriptions and representations, cybersquatting, and unfair competition.[248] The jury ultimately found trademark infringement.[249] In a similar case in Italy, the *Tribunal Ordinario di Roma* ruled against a company that had created and sold NFTs portraying a famous soccer player posing in his club's jersey. While the company had received approval from the soccer star, it neglected to acquire permission from the club for displaying the club's trademark. The court granted the club's request for an injunction and found trademark infringement against the defendant.[250]

The issues raised in the aforementioned disputes are linked to various policy considerations that have been complicated by the rapidly developing NFT market. The mechanical application of IP laws hardly provides a satisfactory solution to these complexities. In the following sections, we illustrate how a dynamic application of IP rules, addressing changing policy considerations, allows for an NFT governance model that comprehensively balances the diverse interests of stakeholders and the public. This model can guide courts, regulators, and market participants as they navigate the evolving NFT market.

## 5.3 | Regulating the NFT market with dynamic IP laws

IP laws are influenced by a range of policy considerations with the overarching objective of striking a balance between public and private interests. More specifically, copyright law is crafted to incentivize creative endeavors while simultaneously promoting the dissemination of knowledge for the benefit of the public.[251] Trademark law, in contrast, focuses on safeguarding consumer welfare and fostering fair competition.[252] In broader terms, IP frameworks strive to minimize transaction costs associated with innovation and contribute to overall economic development.[253] The following sections analyze various public interests that influence IP policy generally and how they can help balancing the interests of multiple stakeholders in the NFT market through dynamically developing the application of IP laws.

---

[240]Koko, *Make Beautiful NFTs in Minutes with These AI Art Generators*, NFT Evening (July 18, 2023), https://nftevening.com/make-beautiful-nfts-in-minutes-with-these-ai-art-generators/.
[241]*See, e.g.*, Complaint, Miramax, LLC v. Tarantino, No. 2:21-cv-08979-FMO-JC (C.D. Cal. Nov. 16, 2021). *Cf. See* Henry Hansmann & Marina Santilli, *Authors' and Artists' Moral Rights: A Comparative Legal and Economic Analysis*, 26 J. Legal Stud. 95, 114 (1997) ("Although painters, sculptors, and other visual artists can retain copyright in their works even after they have sold the original, that copyright covers principally reproductions; it gives the artist much less control over the uses made of his original painting or sculpture once that object is sold by the artist.").
[242]*TamarindArt*, No. 1:22-cv-00595, 2022 WL 195808 (S.D.N.Y. Jan. 21, 2022).
[243]*Id.* at *7.
[244]*Id.* at *8–9.
[245]Complaint, Hermès Int'l v. Rothschild, No. 22-cv-384, 21, 24–25 (S.D.N.Y. Jan. 14, 2022).
[246]*Id.* at *14.
[247]*Id.* at *21, 24–25.
[248]*Id.* at *34–42.

[249]Hermès Int'l v. Rothschild, No. 22-cv-384, 2023 U.S. Dist. LEXIS 109010, at *1 (S.D.N.Y. Jun. 23, 2023). During the pendency of the appeal filed by Rothschild, the court refused to accept MetaBirkins smart contracts and NFTs as collateral to secure the judgment in lieu of a bond regardless of the problems of trademark dilution or infringement. Order, Hermès Int'l v. Rothschild, No. 22-cv-384, 3 (S.D.N.Y. Dec. 29, 2023).
[250]*See* Monia Baccarelli & Marco Lonero Baldassarra, *Italy: Juventus FC Scores Landmark Win for a TM Infringement Case in the Metaverse*, Mondaq (Nov. 29, 2022), https://www.mondaq.com/italy/trademark/1255084/juventus-fc-scores-landmark-win-for-a-tm-infringement-case-in-the-metaverse.
[251]*See, e.g.*, Jyh-An Lee, *The Greenpeace of Cultural Environmentalism*, 16 Widener L. Rev. 1, 25 (2010); L. Ray Patterson & Craig Joyce, *Copyright in 1791: An Essay Concerning the Founders' View of the Copyright Power Granted to Congress in Article I, Section 8, Clause 8 of the U.S. Constitution*, 52 Emory L.J. 909 (2003).
[252]*See* Wang, *supra* note 23, at 380–81; Mark P. McKenna, *The Normative Foundations of Trademark Law*, 82 Notre Dame L. Rev. 1839, 1841, 1848 (2007). Stacey L. Dogan & Mark A. Lemley, *Trademarks and Consumer Search Costs on the Internet*, 41 Hous. L. Rev. 777, 778 (2004).
[253]Richard A. Posner, *Intellectual Property: The Law and Economics Approach*, 19 J. Econ. Persp. 57, 62 (2005). *See generally* Andrew A. Toole et al., Intellectual Property and the U.S. Economy (3rd ed. 2022). *Cf.* Mark A. Lemley, *Taking the Regulatory Nature of IP Seriously*, 92 Tex. L. Rev. 107, 107 (2014) (explicating the policy nature of IP laws, which govern the market entry and prices).

### 5.3.1 | Balanced incentive policy fostering innovation and creativity

Incentivizing innovation and creativity has always been a major policy goal in IP legislation.[254] It is largely achieved by protecting creators' economic interests while simultaneously making their creations available to the public. However, in practice, IP laws occasionally undermine this goal by driving up the price of knowledge, thereby stifling innovation and creativity.[255] For example, to maximize their profits, copyright holders tend to pursue greater control over their works,[256] which leads to limited access to knowledge and innovation sources.[257] Similarly, NFTs are designed to advance creators' economic interests to promote creativity and innovation in the burgeoning NFT market.[258] However, this goal may be undermined when the creator and the copyright holder are not identical, and the licensing agreement between them provides exclusive rights to the copyright holder.[259] In such cases, creators may be disappointed in their expectation to be able to mint NFTs from their creations or derivative works after they have assigned or exclusively licensed the underlying copyright. To solve disputes between creators and copyright holders, we argue that policy choices related to the overarching goals of innovation and creativity can help determine the appropriate scope of conflicting IP interests and balance stakeholder interests. The following subsections provide an illustration for such a case and discuss the proper allocation of copyright interests between stakeholders in two scenarios.

*Balance between creators, copyright holders, and licensees*
Twenty-eight years after Quentin Tarantino wrote and directed the award-winning screenplay Pulp Fiction, the film studio Miramax sued him for his derivative works in the NFT market.[260] In 1993, Tarantino executed an agreement that assigned "all rights (including all copyrights and trademarks) in and to [Pulp Fiction] . . . now or hereafter known including without limitation the right to distribute the Film in all media now or hereafter known" to Miramax.[261] In this assignment, he reserved the rights to "soundtrack album, music publishing, live performance, print publication (including without limitation screenplay publication, 'making of' books, comic books and novelization, in audio and electronic formats as well, as applicable), interactive media, theatrical and television sequel, and remake rights, and television series and spinoff rights."[262] In 2021, Tarantino minted NFTs for uncut scenes, handwritten scripts, and exclusive custom commentary on the film and announced an auction of the NFTs.[263] Miramax subsequently sued Tarantino for breach of contract, copyright, trademark infringement, and unfair competition.[264]

Miramax claimed that the intended sale of original script pages and unpublished scenes as NFTs was a one-time transaction rather than "print publication" or "screenplay publication" and therefore not covered as an authorized use by their 1993 agreement.[265] Tarantino contended that he was acting within his reserved rights.[266] While Miramax and Tarantino reached a settlement in September 2022,[267] the case exemplifies important issues concerning the appropriate allocation of IP interests, taking the aforementioned policy considerations into account.

According to general contract doctrine, courts must take into consideration not only the literal meaning of terms but also the parties' intentions and prevailing business practices at the time of signing, when interpreting licensing terms in dispute.[268] The terms used to designate Miramax's acquired rights, such as "all," "now and hereafter known," and "without limitation,"[269] were relatively broad, general, and forward-looking. Miramax argued that Tarantino's expressly reserved rights were merely an exception to Miramax's broad "catch-all" rights[270] and therefore any right that was not explicitly reserved for Tarantino would remain with the assignee, Miramax. Contrastingly, Tarantino argued that at the time of signing the agreement with Miramax, the concept of NFT did not exist at all.[271] Therefore, it was not the parties' intention to include this technology and its related activities in the license agreement.[272]

---

[254]See, e.g., Sara K. Stadler, *Incentive and Expectation in Copyright*, 58 HASTINGS L.J. 433, 436–36 (2007) (taking copyright as an example and reminding the history of extending the length of copyright protection).
[255]See e.g., Mark A. Lemley, *IP in a World Without Scarcity*, 90 N.Y.U. L. REV. 460, 462–63 (2015).
[256]*Id.* at 497.
[257]See, e.g., JYH-AN LEE, NONPROFIT ORGANIZATIONS AND THE INTELLECTUAL COMMONS 66–67 (2012) (explaining how commercial publishers use copyright and technology to limit researchers' access to knowledge).
[258]*See supra* text accompanying notes 133–136.
[259]*See supra* section 2.5.
[260]Complaint, Miramax, LLC v. Tarantino, No. 2:21-cv-08979-FMO-JC (C.D. Cal. Nov. 16, 2021).
[261]*Id.* at 4–5.
[262]*Id.* at 5.
[263]*Id.* at 1–2.
[264]*Id.* at 7. Further discussions over the legal issues in this case are addressed in the next Section. *See infra* section 5.3.1.
[265]Kal Raustiala & Chris Sprigman, *Guest Column: Tarantino* vs. *Miramax — Behind the NFT 'Pulp Fiction' Case, and Who Holds the Advantage*, HOLLYWOOD REP. (Nov. 24, 2021), https://www.hollywoodreporter.com/business/digital/tarantino-miramax-pulp-fiction-nft-1235052378/.
[266]Bryan Freedman, *Quentin Tarantino's Lawyer Bryan Freedman Says He Has Right to Sell 'Pulp Fiction' NFTs*, BRYANFREEDMAN (Nov. 17, 2021), https://bryanfreedman.net/quentin-tarantinos-lawyer-bryan-freedman-says-he-has-right-to-sell-pulp-fiction-nfts/. *See also* Reply, Miramax, LLC v. Tarantino, No. 2:21-cv-08979-FMO-JC 4, 6–10 (C.D. Cal. July 7, 2022) ("[Reserved rights for the author] encompassed every aspect of the rights granted in the [novel] under the Copyright Act except the right to distribute this particular completed Film based on the [novel]. The fact that the [novel] was a required 'element' of the Film is irrelevant to the contractual reality that all of the rights in the [novel] other than the right to distribute the Film based on the [novel] remained expressly reserved . . . If all rights in the [novel] had been transferred to Miramax, any alteration to the [novel] would have constituted the creation of a derivative work, and [the author and the film producer] would have had to have sought and obtained Miramax's full permission to make that change. No such permission was required because no rights in the [novel] (aside from the right to distribute the Film) were assigned . . . [The] 'now or hereafter know' [rights granted to Miramax were] expressly subject to Mr. Tarantino's reserved rights, which included all publication rights in his screenplay—necessarily including rights in technologies to be developed in the future.").
[267]Notice of Settlement, Miramax, LLC v. Tarantino, No. 2:21-cv-08979-FMO-JC (C.D. Cal. Sept. 8, 2022).
[268]*See* Random House, Inc. v. Rosetta Books, LLC, 283 F.3d 490, 492 (2d Cir. 2002) (ruling that when determining whether a copyright license extends to a new form of use, a court should consider the "customs, practices, usages and terminology as *generally* understood in the . . . trade or business at the time of contracting").
[269]Complaint, Miramax, LLC v. Tarantino, No. 2:21-cv-08979-FMO-JC 4–5 (C.D. Cal. Nov. 16, 2021).
[270]*Id.* at 5.
[271]Caroline Rimmer, *Miramax v. Tarantino: Pulp Fiction and NFTs*, COLUM. J.L. & ARTS (Dec. 1, 2021), https://journals.library.columbia.edu/index.php/lawandarts/announcement/view/461#_ftn4.
[272]*Id.*

Accordingly, Miramax's argument would have been more convincing if it had considered technological development and IP policy considerations beyond textual interpretation.[273] In this regard, we argue it is a reasonable and fair assumption that copyright assignees, rather than original authors, should hold the right to mint NFTs when copyright assignment agreements are silent on NFT-related activities. In cases such as *Miramax v. Tarantino*, disputes arise from the deployment of technologies, particularly blockchains, that were developed long after the copyright agreement was signed. In determining whether a party has the right to mint NFTs through blockchains, courts need to consider whether the party is depriving the other party's interest, as expressed in the original assignment agreement.[274] In doing so, the court may consider whether the original author could have agreed on the same price for a copyright assignment that included the right to mint NFTs when the parties closed the transaction.[275] It is reasonable to assume that the original author would be constrained from minting an NFT without authorization from the copyright holder as double minting decreases the value of not only the NFT but also its associated copyrighted work.[276] One possibility for an original author to revive his or her right to mint NFTs from the copyright assignee is to exercise a reversionary right.[277] This right allows the author or his or her heirs to terminate a license or assignment agreement and regain full copyrights or renegotiate the terms of the agreement with the licensee or assignee after a certain period of time has passed since the agreement was signed.[278] Although reversionary rights have rarely been exercised,[279] the growing NFT market provides an opportunity for policymakers to re-evaluate the justification of this controversial right to foster creativity.[280]

A similar debate may occur between a copyright holder and a licensee when the license is silent on the right to mint NFTs. In this situation, static IP laws and licensing rules are clear enough to resolve the issue: a licensee does not have the right to mint an NFT from the digital assets covered by IP unless the license agreement explicitly allows him or her to do so.[281] Hence, agreeing to a license rather than an assignment agreement does not cause an author to lose his or her exclusive right to mint NFTs.[282]

*Balance between copyright holders and unauthorized follow-up creators*

A critical contribution of NFTs to innovation and creativity is that they attract creators other than the authors of original works to build follow-up works.[283] BAYC's open licensing model has attracted numerous buyers to design derivative works of generic apes (e.g., MAYC).[284] The CC0 license, used by an increasing number of NFT projects, encourages artists to donate their digital artworks to the public domain, facilitating further creative activities and relevant NFT minting.[285] Nevertheless, without a license or clear authorization, whether NFT buyers can legally create and use follow-up works based on the works embedded in NFTs is still uncertain.

This question revives a long-term policy debate about the treatment of derivative works in copyright law that has existed prior to the advent of NFTs.[286] Derivative work rights refer to the copyright holder's exclusive right to prepare additional works based on preexisting works under copyright protection by recasting, reforming, or adapting them.[287] The nature of derivative work rights indicates the scope of copyright, which then determines whether other parties can legally produce follow-up works without authorization from the copyright holders. Some scholars argue that releasing the exclusive right from copyright holders to the public can spur creativity, as it encourages others to create derivative works and thus leads to a "cultural conversation."[288] By contrast, others contend that strengthening the

---

[273]*But see* Dieli, *supra* note 18, at 7–8 (believing that contract law can handle the interpretation well and rejecting NFTs as new entities).

[274]*See, e.g.*, Kirke La Shelle Co. v. Paul Armstrong Co., 263 N.Y. 79 (N.Y. 1933) (ruling that a party cannot deprive the other party's interest when applying new technologies not addressed in contracts).

[275]Some commentators argue that, based on the general market performance of NFTs by 2023, it is reasonable to assume that most original authors would agree that previous copyright assignments included the transfer of rights associated with NFT related activities. *See* Daniel Ruby, *36+ Interesting NFT Statistics 2023 [Trends and Infographics]*, DEMANDSAGE (Feb. 24, 2023), https://www.demandsage.com/nft-statistics/ ("According to NFT statistics, the majority of primary sales were for less than 100 USD."). *But see* David Vaver, *Authors' Moral Rights and the Copyright Law Review Committee's Report: W(h)ither Such Rights Now?*, 14 MONASH U. L. REV. 284, 289 (1988) (arguing that there is always imbalanced bargaining power between authors and copyright acquires).

[276]*See, e.g.*, Miller v. Universal Pictures Co., 10 N.Y.2d 972, 982 (N.Y. 1961) ("[D]iminishes and possibly destroys the value of plaintiff's stake in the record field by means of a violation of the contract.").

[277]*See* Music Sales Corp. v. Morris, 73 F.Supp. 2d 364, 372 (S.D.N.Y. 1999); 17 U.S.C. §§ 203 (a)(3), 304(c).

[278]17 U.S.C. §§ 203(2), 203(4); *see* Robert S. Meitus, *Commentary: Revisiting the Derivative Works Exception of the Copyright Act Thirty Years After Mills Music*, 5 IP THEORY 60, 61 (2015).

[279]Joshua Yuvaraj et al., *U.S. Copyright Termination Notices 1977–2020: Introducing New Datasets*, 19 J. EMPIRICAL STUD. 250 (2022).

[280]*See* Martin Kretschmer, *Copyright Term Reversion and the "Use It or Lose It" Principle*, 1 INT'L J. MUSIC BUS. RES. 44, 44 (2012) ("Term reversion could be a key tool for opening up unexploited back-catalogues, and enabling artist-led cultural and social innovation."). It should be noted that the statutory requirements for revisionary right vary across jurisdictions. *See, e.g.*, PAUL HEALD, COPYRIGHT REVERSION TO AUTHORS (AND THE ROSETTA EFFECT): AN EMPIRICAL STUDY OF REAPPEARING BOOKS 48–60 (2017), http://www.serci.org/congress_documents/2018/heald.pdf (reviewing the rules of reversionary rights around the world). For example, although the United Kingdom abolished reversionary rights, some other commonwealth jurisdictions, such as Canada, still retain the reversionary rights for authors. *See* Copyright Act (1956) § 50, 8th Schedule 6 (U.K.). *Id.*

[281]*See supra* section 6.1. *See also* Davis v. Blige, 505 F.3d 90, 101 (2d Cir. 2007).

[282]*Cf.* Jason Mazzone, *Copyright Easements*, 50 AKRON L. REV. 725, 737–38 (2016) (advising artists to assign or license narrow rights while retaining some rights to exploit future benefits from their works).

[283]*See supra* section 2.5.

[284]While there have been only 10,000 images of bored apes sold on OpenSea, there are 20,000 mutant apes designed from the generic apes. Users are allowed to create new apes. Meanwhile, AI is also used to design unlimited apes. *See, e.g.*, Zeynep Geylan, *The Creators of Bored Ape Yacht Club Acquires the Rights to CryptoPunks and Meebits*, CRYPTOSLATE (Mar. 12, 2022), https://cryptoslate.com/the-creators-of-bored-ape-yacht-club-acquires-the-rights-to-cryptopunks-and-meebits/; *This AI Generates Bored Apes That Are Unique, Free—and Fungible*, GREAT OFFICIALS (May 6, 2022), https://gadgetofficials.com/this-ai-generates-bored-apes-that-are-unique-free-and-fungible/; *This A.I. Creates Infinite NFTs*, MARTECH-VIDEOS ONDECKEO (May 7, 2022), https://martech-videos.ondeckeo.com/this-a-i-creates-infinite-nfts-video/.

[285]*See, e.g.*, Eric James Beyer, *CC0 and NFTs: Understanding Ownership*, NFT NOW (May 5, 2022), https://nftnow.com/features/cc0-and-nfts-understanding-ownership/. Flashrekt & Scott Duke Kominers, *Why NFT Creators Are Going cc0*, A16ZCRPTO (Aug. 3, 2022), https://a16zcrypto.com/cc0-nft-creative-commons-zero-license-rights/ (listing examples of creators that have adopted the CC0 license, such as A Common Place, Anonymice, Blitmap, Chain Runners, Cryptoadz, CryptoTeddies, Goblintown, Gradis, Loot, mfers, Mirakai, Shields, and Terrarium Club).

[286]*See* Pam Samuelson & Members of the Copyright Principles Project, *The Copyright Principles Project: Directions for Reform*, 25 BERKELEY TECH. L.J. 1175, 1176–79 (2010).

[287]*See* 17 U.S.C. §§ 101, 106(2).

[288]Rebecca Tushnet, *Free to Be You and Me? Copyright and Constraint*, 128 HARV. L. REV. F. 125, 126 (2015). *See also e.g.*, Linford, *supra* note 116, at 193–94.

rights of copyright holder to prepare derivative work can encourage knowledge production by providing enhanced protection and incentivizing the copyright holder to create "more."[289] In the U.S. copyright regime, the support for derivative work rights of original authors is inconsistent, leading to varying risks of copyright infringement for NFT minters and buyers on a case-by-case basis. The overarching trend indicates an expansion of derivative work rights.[290] Given this trajectory, courts may very likely hold that NFTs qualify as derivative works, necessitating authorization from copyright holders.[291] However, in certain instances, courts have rejected such claims, allowing unauthorized authors to create follow-up works under defenses like fair use or freedom of expression.[292] This approach, aimed at moderating copyright protection, could potentially foster the emergence of derivative works in conjunction with NFT transactions.[293]

In his dispute with Miramax, Tarantino argued that his use of the works (i.e., minting NFTs) was based on fair use, freedom of speech, and the clean-hands doctrine.[294] However, these defenses entail significant uncertainty. First, despite calls to expand fair use in the virtual world to foster creativity,[295] no court has thus far recognized NFT minting as a noninfringing fair use.[296] This may be attributed to copyright holders increasingly perceiving that NFT-related activities have significantly impacted the market value of copyrighted works. Second, for a First Amendment defense to succeed, the best strategy is to characterize NFTs as parodies or "criticism and comment."[297] However, such a defense is mainly applied to political commentary, a category to which *Pulp Fiction* and many other NFTs do not belong.[298] Third, although it has remained unclear how Tarantino's defense of Miramax's "unclean hands" would unfold, even if it was successful, such a defense in equity could only exonerate Tarantino from equitable liabilities, such as an injunctive relief, but would not shield him from statutory liabilities like damages.

Copyright holders who are also original authors may further secure control over their work through moral rights.[299] Innovation theories arguably support moral rights as an inspirational motivation to promote creation and innovation.[300] This is primarily because moral rights constrain the derivative works generated by subsequent authors.[301] However, moral rights may be less important than derivative work rights in NFT transactions. First, authors creating NFT artworks may prioritize economic interests over moral rights.[302] Blockchains and smart contracts provide a more efficient and effective means of securing royalties than enforcing their moral rights.[303] Second, moral rights are limited in U.S. copyright law.[304] While civil law jurisdictions place a greater emphasis on moral rights for authors' personal and dignity concerns than the United States,[305] the scope and strength of protection are inconsistent.[306] It is improbable that moral rights will become the primary policy for innovation, and NFT-related activities are no exception.[307]

### 5.3.2 | Fair competition and consumer protection

Unlike copyright law, trademark law is designed to eradicate unfair competition and prevent competitors from confusing consumers by free riding on others' brands.[308] Nevertheless, similar to copyright

---

[289]See, e.g., Deidrè A. Keller, *Recognizing the Derivative Works Right as a Moral Right: A Case Comparison and Proposal*, 63 Case W. Rsrv. L. Rev. 511, (2012); Julie A. Hopkins & Jordan Kuchta, *NFTs and Intellectual Property: What You Need to Know*, Md. St. Bar Ass'n (Sept. 21, 2022), https://www.msba.org/nfts-and-intellectual-property-what-you-need-to-know/. *But see* Tushnet, *supra* note 288, at 125 (questioning that more by the same author always means better).

[290]Daniel J. Gervais, *The Derivative Right, or Why Copyright Law Protects Foxes Better than Hedgehogs*, 15 Vand. J. Ent. & Tech. L. 785, 794 (2013).

[291]See Blanch v. Koons, 467 F.3d 244, 252 n.4 (2d Cir. 2006) ("A derivative use can certainly be complementary to, or fulfill a different function from, the original.").

[292]17 U.S.C. §107. *See, e.g.*, Twin Peaks Prods. v. Publ'ns Int'l, Ltd., 996 F.2d 1366, 1375–76 (2d Cir. 1993) (suggesting that transformative use is fair use and is exempted from authorization from copyright holders).

[293]Joseph P. Fishman, *Creating Around Copyright*, 128 Harv. L. Rev. 1333, 1336–37 (2015).

[294]See Quentin Tarantino's and Visiona Romantica Inc.'s Answer to the Complaint, Miramax, LLC v. Tarantino, No. 2:21-cv-08979-FMO-JC 15–16 (C.D. Cal. Dec. 9, 2021).

[295]See, e.g., Matthew R. Farley, *Making Virtual Copyright Work*, 41 Golden Gate U. L. Rev. 1, 22–25 (2010).

[296]Cohen et al., *supra* note 26.

[297]Conrad, *supra* note 26, at 142.

[298]Id.

[299]Moral rights protect the non-pecuniary rights of authors, such as the right of integrity, and rights to prevent distortion and alternation. 17 U.S.C. § 106 A; 17 U.S.C. § 1202(b).

[300]See generally Roberta R. Kwall, *Inspiration and Innovation: The Intrinsic Dimension of the Artistic Soul*, 81 Notre Dame L. Rev. 1945 (2006) (highlighting the intrinsic role of moral rights in promoting innovation); Runhua Wang, *Modify State "Piracy" After Allen: Introducing Apology to the U.S. Copyright Regime*, 69 Buff. L. Rev. 485, 544–48 (2021) (discussing the importance of spiritual motivation to innovation); Harry First, *Controlling the Intellectual Property Grab: Protect Innovation, Not Innovators*, 38 Rutgers L.J. 365, 372 (2007); Lior Zemer, *Moral Rights: Limited Edition*, 91 B.U. L. Rev. 1519, 1522 (2011); Jane C. Ginsburg, *The Most Moral of Rights: The Right to be Recognized as the Author of One's Work*, 8 Geo. Mason J. Int'l Comm. L. 44 (2016) (criticizing the weak protection of attribution rights); Christopher Jon Sprigman et al., *What's a Name Worth?: Experimental Tests of the Value of Attribution in Intellectual Property*, 93 B.U. L. Rev. 1389 (2013) (arguing attribution rights as meaningless).

[301]See, e.g., Teter v. Glass Onion, Inc., 723 F. Supp. 2d 1138, 1146 (W.D. Mo. 2010) (finding copyright infringement in an unauthorized derivative work and opposing the claim of moral rights protection for unqualified subject matter); Jacobsen v. Katzer, 535 F.3d 1373, 1380 (Fed. Cir. 2008) (reminding the legal risks of copyright infringement and breach of contract against unauthorized modification of codes); Wrench LLC v. Taco Bell Corp., 256 F.3d 446, 456–59 (6th Cir. 2001) (holding that breach of implied-in-fact contract claim is not preempted by the Copyright Act).

[302]See Adam Iscoe, *Burnt Banksy's Inflammatory N.F.T. Not-Art*, New Yorker (May 10, 2021), https://www.newyorker.com/magazine/2021/05/17/burnt-banksys-inflammatory-nft-not-art; *cf.* Michael D. Murray, *NFTs Rescue Resale Royalties? The Wonderfully Complicated Ability of NFT Smart Contracts to Allow Resale Royalty Rights*, 14 J.L. Tech & Internet 206 (July 27, 2022). *But see* Alex Swanson, *Will—and Should—VARA Cover NFTs?*, J. Intell. Prop. & Ent. L (Nov. 30, 2021), https://blog.jipel.law.nyu.edu/2021/11/will-and-should-vara-cover-nfts/ (raising concerns that burning NFTs may trigger the right of integrity, which is a moral right).

[303]See Alex Gomez, *NFT Royalties: What Are They and How Do They Work?*, Cyber Scrilla, https://cyberscrilla.com/nft-royalties-what-are-they-and-how-do-they-work/ (last visited Aug. 2, 2022) (suggesting the perpetuity of a smart contract).

[304]See, e.g., Fahmy v. Jay-Z, 891 F.3d 823, 831 (9th Cir. 2018) ("The Copyright Act recognizes some moral rights, but only for certain 'work[s] of visual art.'").

[305]See, e.g., Jean-Luc Piotraut, *An Authors' Rights-Based Copyright Law: The Fairness and Morality of French and American Law Compared*, 24 Cardozo Arts & Ent. L.J. 549, 595 (2006) (comparing moral rights between the U.S. and France).

[306]See Charles A. Marvin, *Author's Status in the United Kingdom and France: Common Law and the Moral Right Doctrine*, 20 Int'l & Comp. L.Q. 675, 676 (1971) ("No common accord has been reached among States which recognize the moral right as to what all of the elements of the bond between the personality of the author and his work are.").

[307]Brandi L. Holland, *Moral Rights Protection in the United States and the Effect of the Family Entertainment and Copyright Act of 2005 on U.S. International Obligations*, 39 Vand. J. Transnat'l L. 217, 230 (2006) (suggesting pecuniary rights can effectively promote innovation and creativity from a utilitarian viewpoint). *See, e.g.*, Greg R. Vetter, *The Collaborative Integrity of Open-Source Software*, 2004 Utah L. Rev. 563, 577 (reminding moral rights concerns when open-source collaborative software was developing).

[308]See, e.g., William M. Landes & Richard A. Posner, The Economic Structure of Intellectual Property Law 166–68 (Harvard Univ. Press 2003). *See also* Mark P. McKenna, *The Normative Foundations of Trademark Law*, 82 Notre Dame L. Rev. 1839, 1900 (2007) (reminding the conventional goal of trademark protection is to reduce consumer search cost and prevent them from being deceived); Robert G. Bone, *Rights and Remedies in Trademark Law: The Curious Distinction between Trademark Infringement and Unfair Competition*, 98 Tex. L. Rev. 1187, 1187 (2020) (stating that trademark infringement and unfair competition share the same underlying policies and the same liability principles).

law, trademark law has been at the center of a great deal of litigation related to NFT minting. This is because NFTs have recently gained strategical relevance for businesses that establish brands and stores in metaverses. Notably, well-established brands such as Amazon, Burberry, McDonald's, Nike, Versace, Vogue, and Walmart have witnessed a surge in trademark applications and the introduction of virtual stores associated with NFTs.[309] As a result, legal questions have arisen regarding the necessity of a valid trademark to launch an NFT and, if so, whether launching an NFT that displays marks owned by others constitutes trademark infringement, akin to the physical world.

In addition to Hermès Int'l v. Rothschild mentioned above,[310] another notable case concerning trademark and NFTs involves Yuga Labs, Inc (Yuga Labs), the creator of the most successful NFT collection in the world—BAYC. Yuga Labs filed a lawsuit against the visual artist Ryder Ripps because his RR/BAYC NFTs were double-minted from BAYC's images on the blockchain.[311] Instead of claiming copyright infringement, Yuga Labs sued Ripps for federal and common law trademark infringement alleging the artist had used BAYC's marks in marketing his RR/BAYC NFTs.[312] Ripps invoked fair use and his First Amendment rights, defending that his use of the BAYC images constituted a form of "appropriation art," aiming at drawing attention to Yuga Labs' racist, neo-Nazi, and alt-right contents and creating social pressure for Yuga Labs to take responsibility for its actions.[313] The court nevertheless rejected Ripps' contention that the RR/BAYC images "express an idea or point of view." Moreover, the court determined that the listings of the RR/BAYC images on NFT marketplaces "contain no artistic expression or critical commentary" and, therefore, could not be protected by the First Amendment.[314] Ultimately, the court sided with Yuga Labs on the trademark infringement claim.[315]

These cases underscore the distinction between trademark law, including protection against trademark dilution, and copyright law. While parties can sometimes justify their unauthorized use of copyrighted works by invoking the First Amendment and fair use doctrine, employing similar defenses in trademark law is a more challenging task. This difficulty arises from different policy considerations underlying trademark and copyright law. While copyright law primarily aims to stimulate creativity, trademark law seeks to foster fair competition[316] and consumer protection.[317] These goals, as the previously mentioned trademark cases illustrate, typically work in favor of the trademark owner.[318] Given its unique policy functions, trademark law presents a double-edged sword for NFT buyers. On one hand, buyers seeking NFTs authenticated by trademark owners may have successfully reduced their search cost under the umbrella of trademark law. On the other hand, buyers with the intentions to commercially exploit NFTs face higher risks of IP infringement. In the absence of a valid endorsement by trademark owners, NFT buyers may overvalue their NFTs because they misunderstand the connection between trademarks and NFTs.[319] This may lead to consumer confusion, a bedrock principle that underlies much of trademark law. Therefore, unlike copyright law, trademark law provides the NFT market with a policy lever by taking into account consumer protection and fair competition.

### 5.3.3 | NFT marketplaces as gatekeepers for IP

NFT platforms and marketplaces play crucial roles in facilitating NFT transactions. As described earlier, these marketplaces are increasingly aware of potential IP issues stemming from NFT transactions.[320] A growing number of NFT marketplaces require licenses from copyright holders to publish artworks online when the latter mint NFTs, based on the goal of preventing disputes between transacting parties, but also to limit their own risk of being held responsible for secondary IP infringement.

NFT marketplaces in the United States may also mitigate the legal risk arising from their users' transactions by seeking immunity through the Digital Millennium Copyright Act's (DMCA) safe harbor.[321] According to the DMCA, Internet service providers (ISPs) can be exempted from copyright infringement liability for third-party content if they follow stipulated procedures, such as notice-and-takedown.[322] The copyright safe harbor is also common in other jurisdictions, such as the European Union and China.[323]

The effectiveness of the aforementioned safe harbor in defending marketplaces like OpenSea is not as straightforward as some practitioners suggest.[324] While most jurisdictions have developed ISP safe harbor regimes, the specifics regarding the allocation of risks and costs between IP holders, ISPs, and their users vary.[325] Additionally, as previously explained, copyright rules may differ from trademark rules in their underlying considerations, and NFTs sometimes raise issues implicating both. These differences stem from the distinct policy goals behind the application and development of safe harbor rules to IP in general and NFTs specifically, and these goals may vary from one jurisdiction to another and evolve over time.[326]

---

[309]Josh Gerben, *Metaverse Trademarks: A Guide to Notable Filings*, GERBENLAW, https://www.gerbenlaw.com/blog/metaverse-trademarks-a-guide-to-notable-filings/ (last visited June 3, 2022).
[310]Hermes Int'l v. Rothschild, No. 22-cv-384, 2022 U.S. Dist. LEXIS 24376 (S.D.N.Y. Feb. 10, 2022). *See supra* section 6.2.
[311]Yuga Labs, Inc. v. Ripps, No. 2:22-cv-04355-JFW-JEM, at 1–2 (C.D. Cal. Dec. 16, 2022).
[312]*Id.* at 2–3.
[313]Yuga Labs, Inc. v. Ripps, No. CV 22-4355-JFW(JEMx), 2023 U.S. Dist. LEXIS 71336, at *2 (C.D. Cal. Apr. 21, 2023).
[314]*Id.* at 16.
[315]Yuga Labs, Inc. v. Ripps, No. 2:22-cv-04355-JFW-JEM, at 16 (C.D. Cal. Dec. 16, 2022).
[316]LANDES & POSNER, *supra* note 308, at 166.
[317]*See, e.g.*, Jyh-An Lee & Yangzi Li, *The Obscure Consumer in the Chinese Intellectual Property Law*, 42 EURO. INTELL. PROP. REV. 55, 61 (2020).

[318]For example, when determining the case, the court distinguished the interests of authors or designers from trademark protection but concentrated on the interests of consumers, which are also consistent with the competition interests of a strong trademark holder. *See* Hermès Int'l v. Rothschild, No. 22-cv-384, 2022 U.S. Dist. LEXIS 17669, at *19 (S.D.N.Y. Feb. 2, 2022) (refusing to apply the Rogers test when claim that "[i]n certain instances, the public's interest in avoiding competitive exploitation or consumer confusion as to the source of a good outweighs whatever First Amendment concerns may be at stake. The Rogers test incorporates these competing considerations.").
[319]*See* Avinandan Banerjee, *Know Four Reasons Why Some NFTs Are Overvalued*, BLOCKCHAIN COUNCIL (Sept. 27, 2022), https://www.blockchain-council.org/nft/know-four-reasons-why-some-nfts-are-overvalued/ (last visited Aug. 16, 2022) (mentioning factors that affect NFT values, which include famous artists and solid brands behind NFTs).
[320]*See supra* section 5.
[321]17 U.S.C. § 512. *See* Sarah Law, *NFT Marketplaces and the DMCA Safe Harbors*, CHI-KENT J. INTELL. PROP. (Feb. 9, 2022), https://studentorgs.kentlaw.iit.edu/ckjip/nft-marketplaces-and-the-dmca-safe-harbors/#_edn10 ("For a website like OpenSea then, whose interactions with the tokens only seem to be the notice and takedown process, the service provider should not have to worry about this disqualification section at all.").
[322]*Id.*
[323]*See, e.g.*, Directive 2000/31/EC, 2000 O.J., Art. 14 (L178); Dianzi Shangwu Fa (电子商务法) [E-Commerce Law of the People's Republic of China] (promulgated by the Standing Comm. Nat'l People's Cong., Aug. 31, 2018, effective Jan. 1, 2019), art. 41–45 (China).

For example, in China, the Hangzhou Internet Court delivered a groundbreaking decision concerning the copyright safe harbor for NFT platforms on April 22, 2022, ruling that the NFT marketplace in question did not meet the requirements for ISP safe harbor.[327] The court found the NFT marketplace Bigverse liable for hosting an infringing NFT minted by one of its users for several reasons, based on Copyright Law and E-Commerce Law in China.[328] First, the marketplace should have known or had reason to know of the illegitimate source of the NFT artwork and should have taken reasonable measures to prevent infringement. Second, due to the self-executing characteristics of smart contracts, an NFT transaction enabled by a smart contract automatically triggers copyright infringement if the underlying artwork infringes copyright. Third, the marketplace was the most cost-effective gatekeeper for preventing copyright infringement because it controlled the contents of the work and the minting process. All data formed in the process of NFT minting and transactions were stored on Bigverse's server, including that of the underlying infringing digital art. The marketplace technically had the capability to review the copyright status of digital art after it was uploaded but before it was minted by the user. Fourth, the marketplace directly received profits from NFTs sold using its service, including "gas fees" from minters as well as 10% royalties and "gas fees" from NFT buyers.[329] In sum, the court mandated that for NFT marketplaces to be immune from copyright infringement liabilities, they need to conduct due diligence regarding copyright status by checking the original artwork, its copyright registration, noninfringement certificates provided by expert institutes, and other materials proving that NFT minters had copyright or copyright licenses for their minting.[330] The decision was upheld by the appellate court.[331]

While there might be different interpretations of these court opinions, the Chinese courts obviously viewed NFT marketplaces as gatekeepers for the underlying copyright. Consequently, while profiting from NFT minting and transactions by the users of NFT marketplaces, the marketplaces are required to share the enforcement costs borne by copyright holders. Specifically, they are required to exercise a certain level of diligence in monitoring potential copyright infringement by their users.

In the United States, although there have not yet been rulings against NFT marketplaces, courts may lean toward leniency regarding the secondary liability of NFT marketplaces.[332] Under U.S. law, ISPs are typically only held liable when they have actual or constructive knowledge of conducts of direct infringement committed by their users. To demonstrate such knowledge, copyright holders usually are required to send ISPs a notice of infringement.[333] Beyond copyright infringement, some scholars suggest that Rule 11(b) of the Federal Rules of Civil Procedure serves as a versatile safe harbor provision applicable to various types of IP, such as trademarks.[334] While there is ongoing debate about the optimal design of safe harbor regimes, Rule 11(b) may effectively supplement the DMCA safe harbor, in addressing the diverse interests of multiple stakeholders, including IP holders, NFT marketplaces, minters, and buyers.[335]

Different designs of the ISP safe harbor regimes reflect how different jurisdictions allocate costs in IP enforcement among stakeholders. Therefore, safe harbor regimes can be seen as a dynamic IP doctrine aiming to balance the interests of IP owners, online platforms, and their users based on various policy considerations, such as cost-effective IP protection and the development of the Internet industry.[336] It is undeniable that clearly defined safe harbor procedures can offer these stakeholders more legal certainty regarding their online activities. This is the primary legislative purpose of the DMCA.[337] With safe harbor rules in place, ISPs have greater

---

[324]See Xinxi Wangluo Chuanbo Baohu Tiaoli (信息网络传播权保护条例) [Regulation on the Protection of the Right of Communication through Information Network] (promulgated by St. Council, effective Mar. 1, 2013), St. Council Gaz., May 18, 2006, Art. 20 (China).

[325]For example, it is more difficult for ISPs in Europe to obtain safe harbor immunity than their counterparts in the United States. The E.U. courts have held ISPs liable for their users' copyright infringement if these ISPs have knowledge of, or control over, such infringement. See, e.g., Case C-160/15, GS Media v Sanoma, 2016 EU.C. 644, paras. 44–55 (2016) (ruling that providing a hyperlink to copyright infringing contents with knowledge constitutes communications to the public and copyright infringement); Case C-682-18, Peterson v. Google, 2021 EU.C. 503 (2021); Cases C-236/08 & C-237/08, Google France SARL v. Louis Vuitton Malletier S.A., 2010, 2010 EU.C. 59, para. 112–113 (2010); Pamela Samuelson, *Pushing Back on Stricter Copyright ISP Liability Rules*, 27 Mich. Tech. L. Rev. 299, 301–03 (2021) (reviewing the controversial legislative history of the Directive); Council Directive (EU) 2019/790, 2019 O.J. (L130). By contrast, ISPs with a similar level of knowledge in the U.S. can still obtain safe harbor exemption. See, e.g., Viacom Int'l, Inc. v. YouTube, Inc., 676 F.3d 19 (2d Cir. 2012). China's approach is similar to that of the United States. See, e.g., Jyh-An Lee, *Tripartite Perspective on the Copyright-Sharing Economy in China*, 35 Comput. L. & Sec. Rev. 434, 444–46 (2019).

[326]See, e.g., Law, *supra* note 322; Samuelson, *supra* note 325; Lee, *supra* note 325.

[327]Shenzhen Qice Diechu Wenhua Chuangyi Youxian Gongsi Su Hangzhou Yuan Yuzhou Keji Youxian Gongsi (深圳奇策迭出文化创意有限公司诉杭州原与宙科技有限公司) [Shenzhen Qice Diechu Cultural Creativity Co., Ltd. v. Hangzhou Yuanyuzhou Technology Co., Ltd.] (Hangzhou Internet Ct. April 22, 2022) (China).

[328]Id.

[329]"Gas fees" refer to a fixed expense of 33 RMB yuan that minters and consumers on Bigverse paid for every transaction of data, such as minting and NFT transactions. Id.

[330]Id.

[331]Shenzhen Qice Diechu Wenhua Chuangyi Youxian Gongsi Su Hangzhou Yuan Yuzhou Keji Youxian Gongsi (深圳奇策迭出文化创意有限公司诉杭州原与宙科技有限公司) [Shenzhen Qice Diechu Cultural Creativity Co., Ltd. v. Hangzhou Yuanyuzhou Technology Co., Ltd.] (Hangzhou Interm. People Ct. Dec. 30, 2022) (China).

[332]See Kevin J. Hickey, Cong. Rsch. Serv., IF11478, Digital Millennium Copyright Act (DMCA) Safe Harbor Provisions for Online Service Providers: A Legal Overview 2 (2020) (showing survey results in which copyrights holders complains but ISPs appreciate the application of the DMCA safe harbor).

[333]See Lucy H. Holmes, Note, *Making Waves in Statutory Safe Harbors: Reevaluating Internet Service Providers' Liability for Third Party Content and Copyright Infringement*, 7 Roger Williams U. L. Rev. 215, 231–37 (2001) (reviewing the legislative history of DMCA, in which the standards of safe harbor rule was designed as intermediate between the interests of ISPs and copyright holders but are still in favor of the former's position).

[334]See Fed. R. Civ. P. 11(c)(2); Robert G. Bone, *Notice Failure and Defenses in Trademark Law*, 96 B.U. L. Rev. 1245, 1281 (2016). See also Irina D. Manta, *Bearing Down on Trademark Bullies*, 22 Fordham Intell. Prop. Media & Ent. L.J. 853, 860 (2012).

[335]Ripps, minters, and Yuga Labs, both copyright holders and ISPs, argued towards safe harbor claims. Judge Walter saved the question for further discussion. See, e.g., Yuga Labs, Inc. v. Ripps, No. 2:22-cv-04355-JFW-JEM 5–11 (C.D. Cal. Mar 17, 2023).

[336]See generally Edward Lee, *Decoding the DMCA Safe Harbors*, 32 Colum. J.L. & Arts 233 (2009).

[337]S. Rep. No. 105-190, at 2 (1998); H.R. Rep. No. 105-551, at 49–50 (1998).

incentives to implement a notice-and-takedown procedure to support online IP enforcement, which in turn supports protection of IP holder's rights while improving legal certainty for ISPs.[338]

Nevertheless, it should be noted that the DMCA was crafted to provide more certainty, particularly to ISPs, regarding their secondary liability.[339] This is because at the time of legislation, Congress believed that this certainty for ISPs could facilitate "the robust development and worldwide expansion of electronic commerce."[340] If policymakers consider revisiting the ISP safe harbor regime for the NFT market, they need to carefully consider new technologies (e.g., blockchains), the legal risks borne by different market participants, and the efficient allocation of costs among them.[341]

Similar to NFT marketplaces, other participants in the NFT market may want to contemplate implementing their own strategies rather than passively wait for courts and policymakers to decide, or worse, simply disregard IP for NFTs. The following section suggests a potential approach for IP holders to manage their rights proactively.

## 6 | IP MANAGEMENT STRATEGIES IN THE RISING NFT MARKET

While IP holders have actively claimed infringement against unauthorized users in virtual frontiers,[342] they do not necessarily extract profits from litigation.[343] Holding an IP alone does not guarantee its enforceability. In an era of increasing NFT activity that leads to ever growing threats but also opportunities for IP holders' rights to capture the economic value of their rights, IP holders have an interest in actively managing their IP. This section provides IP holders with advice on how to do so through copyright licensing and the use of design patents.

### 6.1 | Tolerance of fan and consumer activities

A 2016 article highlighted that it may not be prudent for IP holders to aggressively enforce their rights against fans and consumers,[344] who, similarly, constitute a substantial portion of NFT participants. There are two reasons for this approach. First, fans and consumers may effectively defend their commercial and noncommercial use based on the fair use doctrine and the First Amendment.[345] Second, IP holders can alternatively view consumers' and fans' usage as part of a strategy to attract attention and expand the market.[346]

Accordingly, instead of soliciting royalties from NFT minting and transactions, some IP holders have launched free NFTs for their existing and potential consumers or fans.[347] Consumer attention has independent value and is a scarce resource,[348] which companies normally pay for through advertisements.[349] Therefore, to attract consumers' attention, IP holders are willing to tolerate the former's free use of IP, such as minting NFTs. As Madhavi Sunder argues, "What is tolerated today may become commodified tomorrow."[350] The NFT community cultivates fans of cryptography and virtual artworks,[351] which can attract potential new buyers and fans.

Some IP holders even donate their works to the public through a CC0 license, allowing for the creation of various derivative works among others.[352] While this norm of sharing has been common in various NFT projects, it may not always be an efficient IP management strategy in the long run. Subsequent entrants that create derivative works may enhance both attention and repetition, and it is costly for IP holders to chase the information regarding the derivative works and compete with subsequent entrants for potential consumers and fans.[353] However, sharing IP with a valid license may still be better than tolerating unauthorized minting and follow-up activities without licenses. The latter approach leaves the NFT market with legal uncertainties and may affect the value of NFTs. A flood of pirated or counterfeit NFTs leads to high authentication costs for consumers and NFT buyers.

We recommend that IP holders consider incorporating NFTs in their licensing activities. For example, they can mint NFTs for their granted IP and deploy smart contracts in the process of signing licenses.[354] NFTs offer an efficient managerial tool to assist in IP licensing and in elevating IP holders and potential licensees to "win-

---

[338]See e.g., Kurt M. Saunders & Gerlinde Berger-Walliser, *The Liability of Online Markets for Counterfeit Goods: A Comparative Analysis of Secondary Trademark Infringement in the United States and Europe*, 32 Nw. J. Int'l L. & Bus. 37, 91 (2011).

[339]S. Rep. No. 105-190, at 1 (1998).

[340]*Id*. at 1–2.

[341]It is noteworthy that on October 5, 2021, Congressman McHenry introduced the Clarity for Digital Tokens Act and called for a "safe harbor" for digital asset start-up projects to further promote these industries. *See* H.R. 5496, 117th Cong. (2021).

[342]Melissa Ung, *Trademark Law and the Repercussions of Virtual Property (IRL)*, 17 Commlaw Conspectus 679, 682–83 (2009).

[343]*See, e.g.*, Amaretto Ranch Breedables v. Ozimals, Inc., No. CV 10-5696, 2013 U.S. Dist. LEXIS 95685, at *4 (N.D. Cal. July 9, 2013); Minsky v. Linden Rsch., No. 1:08-CV-819, 2008 U.S. Dist. LEXIS 143547, at *1 (N.D.N.Y. Sept. 4, 2008).

[344]Jessica M. Kiser, *Brands as Copyright*, 61 Vill. L. Rev. 45, 45–48 (2016) (suggesting that smart brand holders would not sue consumer infringers).

[345]*Id*. at 248–49. *See also* Rebecca Tushnet, Note, *Legal Fictions: Copyright, Fan Fiction, and a New Common Law*, 17 Loy. L.A. Ent. L.J. 651, 681 (1997) (defending fanfiction as fair use); Conrad, *supra* note 26, at 145, 150 (defending the creation of NFTs under the First Amendment); Sunder, *supra* note 119, at 245–56 (exploring the application of fair use defense to safeguard fan activity), at 227 (criticizing a narrow application of fair use for market failure and parody only). *Cf.* Conrad, *supra* note 26, at 227 (rejecting NFT minting to be a fair use of trademarks, even though it may be a fair use of copyrights). *See also* Campbell v. Acuff-Rose Music, Inc., 510 U.S. 569, 579 (1994) (finding a transformative use over a commercial use to be not infringement).

[346]*Id*. (citing B. Joseph Pine II & James H. Gilmore, The Experience Economy ix (updated ed. 2011)).

[347]*See supra* section 2.5.

[348]*See generally* Linford, *supra* note 116.

[349]*Id*. at 146, 168.

[350]Sunder, *supra* note 119, at 228.

[351]*See* Lee, *supra* note 128 (introduced five reasons why people enter the NFT market, the last of which is being advertised by internet celebrities). *See, e.g.*, Fanaply, https://www.fanaply.com/ (last visited Aug. 7, 2022). Fanaply is an NFT platform that encourages fans to mint their favorites of collectibles in the material world.

[352]*See supra* section 5.

[353]*See* Linford, *supra* note 116, at 193–94 (suggesting that consumers receive few creative products with homogeneous works).

[354]*Cf.* Andrew Patty II & Andrew W. Coffman, *Tokenization of IP: Efficient Tool to Manage IP Rights or a Fool's Errand?*, Am. Bar Ass'n (April 3, 2023), https://www.americanbar.org/groups/intellectual_property_law/publications/landslide/2022-23/march-april/tokenization-ip-efficient-tool-manage-ip-rights-fools-errand/ (suggesting "tokenize" IP for reducing transaction costs, broadening the size of potential licensees, and providing more flexibility for nascent or small businesses.).

win" positions.[355] Many IP licensing negotiations are time-consuming and involve uncertainties for several reasons, such as the status of patent applications and the lack of transparency of comparable licenses.[356] These transaction costs can be effectively reduced through the use of smart contracts and NFTs.[357] With NFTs, potential licensees can quickly use art or designs at a set price.

## 6.2 | Design patent as an alternative form of protection

Embarking on an NFT business may not be a viable option for every artist. Not only may they simply lack interest, but the process of designing and managing a new virtual business model can be financially burdensome, and the returns on investment remain anything but certain. For instance, giant IP holder McDonald's, boasting a global fanbase of all ages, introduced its virtual McWorld in 2008,[358] yet yielded returns that remained far below expectations as the McWorld has been largely confined to competitive virtual games. McDonald's failed to align the virtual McWorld with its brand to effectively engage consumers.[359] Given McDonald's lack of success, it is likely to be even more difficult for individual artists and small firms to manage their IP within the NFT network.

An alternative for artists and innovators is to explore design patents as a form of protection. The relationships between inventors, assignees, and licensees of design patents are generally more straightforward when compared to copyright. This is due to the fact that design patents do not entail the legal complexities associated with derivative work rights and reversionary rights.[360] Design patents protect the ornamental aspects of a visual work and must be novel, nonobvious, and original.[361] Some subject matters of design patents are also entitled to copyright protection, such as graphic or architectural works.[362] They may extend protection beyond the scope of copyrights and trademarks, particularly for virtual designs and three-dimensional works.[363] Moreover, the process of securing protection through a design patent is both faster and less arduous than obtaining trade dress protection. The latter typically necessitates the establishment of a secondary meaning, adding a layer of complexity to the process.[364]

Nonetheless, the pursuit of design patents for works minted in NFTs has the drawbacks of high acquisition costs and maintenance fees. Because of the rigorous standards of novelty and nonobviousness requirements, it is sometimes not easy for artists and designers to obtain design patents.[365] Although NFTs may reduce transaction costs in licensing design patents, expensive institutional fees are inevitable.[366]

The value of design patents, however, has been constantly underestimated not limited to but including in the realm of NFTs.[367] While NFTs have been minted from expired design patents to revive the value of these abandoned and forgotten designs,[368] it is uncommon for visual artworks associated with NFTs to be protected by design patents. Though it may take time to fully understand how the mechanism of design patents might work in metaverses,[369] there is significant untapped potential for designs to be minted in NFTs.[370] NFTs that create greater financial and marketing opportunities for artists and designers may enable them to consider using design patents and leveraging their creative works in the future.

## 7 | CONCLUSION

Alluring terms like "scarcity" and "communities" have drawn artists, investors, and consumers into the rapidly expanding NFT market.

---

[355]*See generally* Seyed Mojtaba Hosseini Bamakan et al., *Patents and Intellectual Property Assets as Non-Fungible Tokens; Key Technologies and Challenges*, 12 Sci. Rep. 2178 (2022) (designing an NFT-based patent framework for enhancing the efficient of patent grants, funding, and licensing, etc.).

[356]*See id.*

[357]*See id.*

[358]Justin Olivetti, *The Game Archeologist: McDonalds' McWorld MMO Is a Thing That Existed*, Massively Overpowered (May 23, 2020, 12:00 PM), https://massivelyop.com/2020/05/23/the-game-archaeologist-mcdonalds-mcworld-mmo-is-a-thing-that-existed/.

[359]*See id.*

[360]*See, e.g.*, Sarah Burstein, *Is Design Patent Examination Too Lax?*, 33 Berkeley Tech. L.J. 607, 611 (2018) (mentioning the favor of design patents in both the judicial system and the USPTO and suggesting the stable validity of design patents).

[361]Elizabeth Ferrill et al., *Demystifying NFTs: Intellectual Property Protections with Design Patents*, Reuters (June 1, 2022, 8:34 AM), https://www.reuters.com/legal/legalindustry/demystifying-nfts-intellectual-property-protections-with-design-patents-2022-06-01/.

[362]Gregory R. Mues, *Dual Copyright and Design Patent Protection: Works of Art and Ornamental Designs*, 49 St. John's L. Rev. 543, 546 (1975) ("To the extent that the artwork can be applied to an article of manufacture, imparting to it an original, ornamental design, the proprietor may merit dual protection [of design patents and copyrights]."). *See also* 17 U.S. C. §102.

[363]*See, e.g.*, Meshwerks, Inc. v. Toyota Motor Sales U.S.A., Inc., 528 F.3d 1258, 1269 (10th Cir. 2008) (rejecting copyright protection for digitalized models). *See also* Jason J. Du Mont & Mark D. Janis, *Virtual Designs*, 17 Stan. Tech. L. Rev. 107, 111–12 (2013) (preferring design patents to copyrights and trade dress for protecting virtual designs, arguing that design patents protect graphical user interface ("GUI") designs better than copyrights and trademarks, and introducing that copyright protection for 3D designs was not available in the U.S. copyright law history and trademark law was silent to protect 3D designs); Rachel Stigler, *Ooey GUI: The Messy Protection of Graphical User Interfaces*, 12 Nw. J. Tech. & Intell. Prop. 215, 216 (2014) (suggesting that design patents protect graphical user interface GUI as a whole, which is broader than copyrights that only protect "exact knock-offs of the design"); *see generally* Andrea D'Andrea, *Copyright and Legal Issues Surrounding 3D Data*, in 3D Data Creation to Curation: Community Standards for 3D Data Preservation (Jennifer Moore et al. eds., 2022) (showing the challenges for 3D designs to obtain copyright protection by judicial cases).

[364]Stigler, *supra* note 363, at 236.

[365]There are requirements of utility, novelty, and non-obviousness for receiving a design patent successfully, which may bar many artists from transitioning from copyrights to design patents. *See* Steve W. Ackerman, *Protection of the Design of Useful Articles: Current Inadequacies and Proposed Solutions*, 11 Hofstra L. Rev. 1043, 1053 (1983) (criticizing the strict requirements for designs).

[366]Patenting fees are high in the United States, which has attracted scholars to explore alternative solutions for a long time. *See, e.g.*, Jonathan S. Masur, *Costly Screens and Patent Examination*, 2 J. Legal Analysis 687 (2010); Gaétan de Rassenfosse & Adam B. Jaffe, *Are Patent Fees Effective at Weeding Out Low-quality Patents?* (Motu Working Paper 15-01), https://motu-www.motu.org.nz/wpapers/15_01.pdf.

[367]*See generally* Sarah Burstein, *Costly Designs*, 77 Ohio St. L.J. 107 (2016) (defending the social welfare contributed by protection for design patents); Sarah Burstein, *Moving Beyond the Standard Criticisms of Design Patents*, 17 Stan. Tech. L. Rev. 305, 306–07 (2013).

[368]Brian L. Frye, *Adopt a Design Patent: 1842–1925*, SSRN (Nov. 29, 2021), https://papers.ssrn.com/sol3/papers.cfm?abstract_id=3934074.

[369]*See* Jason J. Du Mont & Mark D. Janis, *The Origins of American Design Patent Protection*, 88 Ind. L.J. 837, 840 (2013) ("One reason why the design patent system has remained largely unexplored in the literature is that scholars have never explained how and why the system came to exist.").

[370]*See* Michael Sun & Jimmy Shen, *Design Patents in the Metaverse*, Lexology (Oct. 28, 2022), https://www.lexology.com/library/detail.aspx?g=9740d1f6-ad36-459e-a573-386aed03c2a2.

However, misconceptions and biases surrounding NFTs and their interplay with IP persist. NFTs, like many emerging technologies, pose intricate challenges to the existing legal framework and have ushered in a dual landscape of opportunities and concerns within society. On one hand, NFTs can serve as a valuable complement to the IP regime, fostering innovation and creativity due to their excludability and rivalry characteristics. On the other hand, the inability of NFTs to definitively establish scarcity for their underlying digital assets has resulted in issues like double minting, batch minting, and over-minting, incurring significant social costs.

We illustrate the value of IP laws in regulating the NFT market and explain why the application of property law or sales law to the NFT market is undesirable. Static and dynamic IP laws can better address the tragedies of the commons and anticommons in the NFT market by considering diverse policy considerations, such as efficient governance of intangible assets, technological development, incentivizing creativity, and promoting fair competition. Moreover, policymakers should give due consideration to the preference of NFT communities for aligning the interests of various stakeholders through IP licenses.

## ACKNOWLEDGEMENTS
We would like to thank Scott Baker, Elena Beier, Gerlinde Berger-Walliser, Alina Ng Boyte, Zachary Catanzaro, George C.C. Chen, Kevin Collins, Danielle D'Onfro, Jens Frankenreiter, Danny Friedmann, Timothy T. Hsieh, Grace Ip, Wei Han, Hui Huang, Yanbei Meng, Jay P. Kesan, Edward Lee, Joseph Lee, Lauren Yu-Hsin Lin, Lisa Macklem, Tyler T. Ochoa, Emma Perot, Lisa Ramsey, Neil Richards, Guy Rub, Sandra Sperino, Brian Tamanaha, David Tan, Peter Yu, and Ronald Yu for their helpful comments. This Article has also benefited from feedback provided in the Intellectual Property Scholar Conference (IPSC) at Stanford Law School, the fourteenth IP Conference: Digital Innovations and Intangible Assets at the Chinese University of Hong Kong Faculty of Law, International Forum on the Future Rule of Law and Digital Law at Renmin University of China, Metaverse Law Conference at the Chinese University of Hong Kong Faculty of Law, Web3 Governance Law and Policy Conference at the University of Manchester, the seventh annual Texas A&M Intellectual Property Scholars Roundtable at Texas A&M University School of Law, and Faculty Workshop at Washington University in St. Louis School of Law.

> **How to cite this article:** Runhua Wang, Jyh-An Lee, Jingwen Liu, 'Unwinding NFTs in the shadow of IP law' (2024) 61 Am Bus Law J 31.

## APPENDIX A

Table A.1 provides an overview of typical IP licensing terms offered by popular NFT projects and marketplaces. It delineates the rights granted to NFT buyers and those retained by NFT minters, who may either be artists or act on behalf of artists when minting NFTs. Notably, some licenses exhibit conflicting clauses or establish contradictory rights between NFT buyers and minters. Additionally, certain terms within these licenses may run afoul of IP laws, such as provisions that forbid NFT minters (i.e., artists) from reserving moral rights. These problematic linguistic constructs or gaps within these licenses underscore the uncertainties in the deployment of IPs by both artists and NFT buyers following NFT transactions.

**TABLE A.1** Sample licensing terms adopted by popular NFT projects and platforms.

| Operators | Stakeholders | Display | Copy | Noncommercial use | Commercial use | Use in trademarks | Derivative works | Moral rights | Raise lawsuits | License type | Governing law |
|---|---|---|---|---|---|---|---|---|---|---|---|
| **Panel 1—NFT Projects** | | | | | | | | | | | |
| World of women | Creators | Y | Y | Y | N | N | N | Y | Y & N | Assignment | France |
| | Buyers | Y | Y | Y | Y | Y | Y | N | Y | | |
| BAYC; MAYC | Creators | N | N | N | Y | N | N | Y | N | Unlimited and exclusive | New York |
| | Buyers | Y | Y | Y | Y | Y | Y | Y | Y | | |
| Vee Friends | Creators | Y | Y | Y | Y | Y | Y | Y | Y | Limited and exclusive | New York |
| | Buyers | N | N | Y | N | N | N | N | N | | |
| Forgotten Runes | Creators | Y | Y | Y | Y & N | Y & N | Y | Y | Y | Unlimited and nonexclusive | California |
| | Buyers | Y | Y | Y | Y & N | Y | Y & N | N | N | | |
| MoonBirds | Creators | Y | Y | Y | Y | Y | Y | Y | Y | Unlimited and nonexclusive | Oregon |
| | Buyers | Y | Y | Y | Y | Y | Y | N | N | | |
| CryptoKitties | Creators | Y | Y | Y | Y | Y | Y | Y | Y | Limited and nonexclusive | Delaware |
| CryptoPunks Meebits | Buyers | Y | Y | Y | Y with a limit | N | N | N | N | | |
| Fanaply | Creators | Y | Y | Y | Y | Y | Y | Y | Y | Limited and nonexclusive | New York |
| | Buyers | Y | N | Y | N | N | N | N | N | | |
| **Panel 2—NFT Marketplaces** | | | | | | | | | | | |
| KnownOrigin | Creators | Y | Y | Y | Y | Y | Y | N | N | Limited and nonexclusive | England and Wales |
| | Buyers | Y | N | Y | N | N | N | N | N | | |
| Foundation | Creators | Y | Y | Y | Y | Y | Y | N | Y | Limited and nonexclusive | California |
| | Buyers | Y | N | Y | N | N | N | N | N | | |
| SuperRare | Creators | Y | Y | Y | Y | Y | Y | N | Y | Limited and nonexclusive | New York |
| | Buyers | Y | N | Y | N | N | N | N | N | | |
| MakersPlace | Creators | Y | Y | Y | Y | Y | Y | N | Y | Limited and nonexclusive | Delaware |
| | Buyers | Y | N | Y | N | N | N | N | N | | |
| La Collection | Creators | Y | Y | Y | Y | Y | Y | Y | Y | Limited and nonexclusive | France |
| | Buyers | Y | N | Y | N | N | N | N | N | | |
| CC0 License | Creators | Y | Y | Y | Y | Y | Y | N | N | Unlimited and nonexclusive | N/A |
| | Buyers | Y | Y | Y | Y | Y | Y | N | N | | |

*Note*: For the NFT projects listed in Panel 1, the operators are deemed NFT minters. They create digital art and NFTs for an artwork or only mint NFTs on behalf of the art's authors. For the NFT marketplaces listed in Panel 2, operators are intermediary platforms facilitating the minting and transacting of NFTs between NFT buyers and creators. Words in italics refer to the rights or title expressed in written language in the licenses. Words in regular font refer to the rights and license types about which the licenses are silent but are deduced according to default rules. The governing law for the default rules is included in the last column. Note: no choice-of-law clause was found in MAYC's terms of service. The terms, nevertheless, include an arbitration clause, with New York as the seat of arbitration. In such an event, arbitrators must determine the applicable law, which is usually decided as the law with which the dispute has the closest connection. Considering the domiciles of the parties, place of execution, place of performance, etc. the law to be applied is most likely to be the law of New York. With regard to Forgotten Runes, the license extends nonexclusive rights for personal use and exclusive rights for commercial use except to NFT minters and licensees in writing, which is de facto nonexclusive. With regard to CC0, note that the license extends nonexclusive rights for personal use and exclusive rights for commercial use except to NFT minters and licensees in writing, which is de facto nonexclusive.